\documentclass{ws-rv9x6}

\usepackage{psfrag}
\usepackage[T1]{fontenc}

\usepackage{amsfonts}
\usepackage{amssymb}
\usepackage{makeidx}
\usepackage{times}
\usepackage{pifont}
\usepackage{bm}
\usepackage{ae,aecompl}
\usepackage{amsmath}
\usepackage{graphicx}

\makeindex

\begin{document}

\setcounter{chapter}{0}

\chapter{SPACETIME APPROACH\index{spacetime approach|(} TO PHASE
  TRANSITIONS}
\markboth{W. Janke and A. M. J. Schakel}{Spacetime Approach to Phase
Transitions}
\author{Wolfhard Janke}
\address{Institut f\"ur Theoretische Physik, Universit\"at Leipzig,
  Augustusplatz 10/11, 04109 Leipzig, Germany}
\author{Adriaan M. J. Schakel} 
\address{Institut f\"ur Theoretische
Physik, Freie Universit\"at Berlin, Arnimallee 14, 14195 Berlin,
Germany}

\begin{abstract}
  In these notes, the application of Feynman's sum-over-paths approach
  to thermal phase transitions is discussed.  The paradigm of such a
  spacetime approach to critical phenomena is provided by the
  high-temperature expansion of spin models.  This expansion, known as
  the hopping expansion in the context of lattice field theory, yields a
  geometric description of the phase transition in these models, with
  the thermal critical exponents being determined by the fractal
  structure of the high-temperature graphs.  The graphs percolate at the
  thermal critical point and can be studied using purely geometrical
  observables known from percolation theory.  Besides the phase
  transition in spin models and in the closely related $\phi^4$ theory,
  other transitions discussed from this perspective include
  Bose-Einstein condensation, and the transitions in the Higgs model
  and the pure U(1) gauge theory.
\end{abstract}

\date{July 11, 2006}

\section{Introduction}
Feynman's spacetime approach to quantum mechanics provides a marvelously
intuitive, yet computationally powerful entry to the quantum
world.\cite{Feynman48} In this sum-over-paths approach, transition
amplitudes are computed by summing over all possible trajectories a
particle can take\index{particle trajectories|(}.  Although it is most
extensively applied to nonrelativistic quantum
mechanics,\cite{KleinertQM} for which it was originally designed, the
worldline\index{worldlines|(} approach also found applications in the
context of relativistic systems.\cite{Feynman50,proptime} (For overviews
and references to a host of applications in high energy physics, see
Refs.~[\refcite{current,Polyakov,Strassler:1992zr,Schubert}].)  In these
notes, we explore its use in the theory of critical
phenomena\index{critical!phenomena|(}.

The possibility to adapt Feynman's sum-over-paths approach to describe
phase transitions derives from particle-field
duality\index{particle-field duality}.  Featuring the particle content
of the system, the spacetime approach provides a completely equivalent
alternative to the quantum field theoretic description.  Although by no
means the only successful method,\cite{FisherRMPII} field theory is
frequently used to describe critical phenomena,\cite{Amit,Dimo,KlSF}
whether the fluctuations are thermal as in phase transitions taking
place at finite temperature, or quantum as in zero-temperature
transitions.  Particle-field duality\index{particle-field duality} then
opens the possibility for a spacetime approach to critical phenomena.
In the context of classical critical phenomena, the line objects
described by the field theory are current lines.

Though not formulated as such, the high-temperature
expansion\index{high-temperature!expansion|(} of spin
models\cite{Stanley,Polyakov} probably provides the first example of the
spacetime approach to critical phenomena (see below).  A particular
intuitive example is offered by the nonlinear O($N$) spin
model\index{O($N$) model} defined on a lattice in the limit $N\to0$,
where the high-temperature graphs\index{high-temperature!graphs|(} form
random self-avoiding walks\index{self-avoiding walks|(}.\cite{deGennes}
In the early 1950ies, Feynman\cite{lambda} applied his approach to the
superfluid phase transition\index{superfluid!phase transition} in liquid
$^4$He, and showed that, in this picture, Bose-Einstein
condensation\index{Bose-Einstein condensation} arises when the
worldlines of individual particles combine to form large lines threading
the entire system.  In the late
1970ies,\cite{BMK,Peskin,TS,StoneThomas,Samuel} the approach was applied
to phase transitions in relativistic lattice models possessing line-like
topological defects, thus mapping these models onto the statistical
mechanics problem of random loop configurations.  The line defects were
pictured as forming a grand canonical ensemble of fluctuation loops of
arbitrary shape and length--a so-called \textit{loop
gas}\index{loop!gas}.  In the 1980ies, the statistical properties of
such loop gases were further investigated in the context of cosmic
strings and their role in phase transitions in the early universe, with
the loop distribution\index{loop!distribution} given a central
position,\cite{Copeland:1988bh,Copetal} as well as in the context of
deconfinement phase transitions\index{confinement}, where percolation
observables were introduced to characterize
them.\cite{Patel:1983sc,Hands:1989cg} More recently, these ideas were
pursued further with the goal to arrive at a complete quantitative
understanding of phase transitions in terms of the geometrical
properties of the loop gas\index{loop!gas} under
consideration.\cite{ABH,NguyenSudbo,loops,Hoveetal,pre01,geoPotts,ht,anomalous}
The resulting geometric description of critical phenomena ties together
various strains of theoretical physics: statistical mechanics, quantum
field theory, polymer physics, and percolation theory\index{percolation
theory|(}--all to be discussed below.

The plan of these notes is as follows.  In Sec.~\ref{sec:lft}, the
particle content of the linear O($N$) symmetric quantum field theory,
defined on a hypercubic lattice, is studied in the spirit of Feynman's
sum-over-paths approach, starting from the noninteracting theory in
Secs.~\ref{sec:noninteracting} and \ref{sec:loopgas}.  Interactions are
perturbatively included in Sec.~\ref{sec:phi4}, where also the spacetime
interpretation of Feynman diagrams\index{Feynman diagrams|(} is pursued.
In Sec.~\ref{sec:spinmodel}, the strong-coupling limit, where the field
theory reduces to the nonlinear O($N$) spin model\index{O($N$) model},
is considered.  The high-temperature
representation\index{high-temperature!representation|(} of the nonlinear
model is shown to provide a purely geometric representation, directly
connected to Feynman's spacetime approach for the linear model.  As a
case study, the high-temperature representation of the Ising
model\index{Ising model|(}, corresponding to $N=1$, on a square lattice
is detailed in Sec.~\ref{sec:IM}.

In Sec.~\ref{sec:CP}, the critical properties of the O($N$) universality
class are examined from the perspective of particle
trajectories\index{particle trajectories} or high-temperature graphs.
In Sec.~\ref{sec:fractal}, their fractal structure is investigated,
while in Sec.~\ref{sec:proliferation}, the proliferation of these lines
is shown to drive the O($N$) phase transition.  In Sec.~\ref{sec:ces},
the fractal structure of the worldlines or high-temperature graphs is
related to the critical behavior.  The general relations are first
applied to self-avoiding random walks, corresponding to the $N\to0$
limit, in two dimensions (Sec.~\ref{sec:saw}), and to arbitrary $-2 \leq
N \leq 2$ (Sec.~\ref{sec:on}) also in two dimensions, where many exact
results are available for comparison.

In Sec.~\ref{sec:mc}, the high-temperature representation of the Ising
model\index{Ising model} is investigated by means of Monte Carlo
simulations\index{Monte Carlo}.  The Metropolis update used
to generate the high-temperature graphs is introduced in
Sec.~\ref{sec:plaq}, while the numerical results are summarized in
Sec.~\ref{sec:num}.

In Sec.~\ref{sec:appl}, the conclusions for the O($N$) universality
class are extended to charged systems (Sec.~\ref{sec:higgs}) and to
Bose-Einstein condensation\index{Bose-Einstein condensation}
(Sec.~\ref{sec:bec}) to demonstrate the generality of the spacetime
approach to phase transitions.

Finally, in Sec.~\ref{sec:dual}, the worldlines or high-temperature
graphs are reinterpreted as line defects in two
(Sec.~\ref{sec:peierls}), three (Sec.~\ref{sec:vortex}), and four
(Sec.~\ref{sec:monopole}) dimensions, respectively, and the so-called
dual field theories\index{dual theory}, which feature these
line-like configurations as topological defects, briefly discussed.

\section{Lattice Field Theory\index{lattice field theory|(}}
\label{sec:lft}
\begin{figure}
\psfrag{x}[t][t][.8][0]{$x_1$}
\psfrag{y}[t][t][.8][0]{$x_2$}
\psfrag{t}[t][t][.8][0]{$\tau$}
\begin{center}
\includegraphics[width=.3\textwidth]{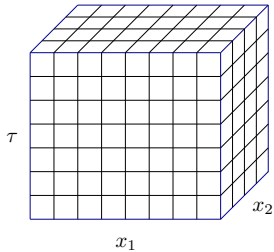}
\end{center}
\caption{Cubic lattice representing three-dimensional discretized
spacetime.
  \label{fig:cubic}}
\end{figure}

Central to our discussion is the O($N$) symmetric relativistic quantum
field theory formulated on a hypercubic lattice in $d$ Euclidean
spacetime dimensions, with the time coordinate analytically continued to
the imaginary axis $\tau$.  As an illustration, the cubic lattice in
Fig.~\ref{fig:cubic} represents discretized spacetime in one time and
two space dimensions $x_1$ and $x_2$.
The theory, describing self-interacting scalar particles of mass $m$, is
specified by the (Euclidean) partition function
\begin{equation} 
\label{Z}
Z = \mathrm{Tr} \, \mathrm{e}^{-S},
\end{equation} 
where  $S$ is the lattice action
\begin{equation} 
\label{Sor}
S = a^d \sum_x \left[ \frac{1}{2a^2} \sum_\mu \left(\varphi_{x+a
      \hat{\mu}} - \varphi_x\right)^2 + \frac{m^2}{2}
      \varphi^2_x + \frac{g}{4!}
      \varphi^4_x \right],
\end{equation} 
with $g$ the coupling constant and $a$ the lattice spacing.  The real
scalar field $\varphi_x$ defined on each lattice site $x$ of the
spacetime box has $N$ components $\varphi_x = \varphi_x^\alpha =
(\varphi^1_x, \varphi^2_x, \cdots , \varphi^N_x)$, with the index
$\alpha=1,2 \cdots, N$ labeling the field components and $\varphi_x^4
\equiv (\varphi_x \cdot \varphi_x)^2$, where the dot product implies a
summation over the field components $\varphi_x \cdot \varphi_x =
\sum_{\alpha=1}^N \varphi_x^\alpha \varphi_x^\alpha$.  Lattice
coordinates, representing discretized spacetime, are specified by $x =
x_\mu= (x_1, x_2, \cdots , x_d)$, with $\mu=1,2,\cdots , d$ and
$x_d=\tau$ denoting the imaginary time coordinate, while $\hat{\mu}$
denotes the unit vector pointing in the $\mu$-direction.  In the
continuum limit $a \to 0$, the lattice action reduces to
\begin{equation} 
\label{Scont}
S = \int \mathrm{d}^d x \left\{ \frac{1}{2} \left[\partial_\mu \varphi(x)
\right]^2 + \frac{m^2}{2} \varphi^2(x) + \frac{g}{4!}  \varphi^4(x)
\right\},
\end{equation}
where $\varphi(x)$ stands for the field defined in continuous spacetime.
Unless otherwise indicated, natural units $\hbar = c = 1$ are adopted
throughout. The trace Tr in Eq.~(\ref{Z}) stands for the sum or integral
over all possible field configurations:
\begin{equation}
\label{Tr} 
\mathrm{Tr} =  \prod_x \int \mathrm{d} \varphi_x.
\end{equation}   
In the continuum limit, where the lattice spacing is taken to zero $a
\to 0$, the right hand defines the \textit{functional} measure denoted
by $\int\mathrm{D} \varphi$.

For numerical simulations, a more convenient form of the lattice action
is obtained by casting Eq.~(\ref{Sor}) in terms of dimensionless fields
and parameters defined via\cite{JanSmit}
\begin{eqnarray} 
\label{K1}
a^{d-2} \varphi^2_x &=& 2 K \, \phi_n^2 \\ 
\label{K2} 
a^{4-d} g &=& 6 \frac{\lambda}{K^2} \\ 
\label{K3}
m^2 a^2  &=& \frac{1-2\lambda N}{K} -2 d.
\end{eqnarray}  
The action then takes the form of an O($N$) spin model\index{O($N$) model}
\begin{equation}
\label{Sorp}
S = - K \sum_{\langle n,n' \rangle} \phi_n \cdot
  \phi_{n'} + \sum_n \phi^2_n + \lambda \sum_n \left(
  \phi_n^2 - N \right)^2.
\end{equation}
Each site on the hypercubic spacetime lattice is now specified by the
vector $n = n_\mu = (n_1, n_2, \cdots , n_d)$ with integer components
$n_\mu = x_\mu/a$ and the sum $\sum_{\langle n,n' \rangle}$ extends over
all nearest neighbor pairs.  In terms of these new dimensionless
variables, the action is independent of the lattice spacing $a$.  The
partition function $Z$ now reads
\begin{equation} 
  Z = \int \mathrm{D} \mu(\phi) \exp \left(K \sum_{\langle n,n'
    \rangle} \phi_n \cdot \phi_{n'} \right),
\end{equation} 
with the on-site measure
\begin{equation}
\label{onsite} 
  \int \mathrm{D} \mu(\phi) = \int \prod_n \mathrm{d} \phi_n \;
      \mathrm{e}^{-\phi^2_n - \lambda \left( \phi_n^2 - N \right)^2}.
\end{equation} 
In the limit $\lambda \to \infty$, the field theory reduces to the
standard O($N$) spin model\index{O($N$) model}, with a ``spin'' variable
$\phi_n$ of fixed length, $\phi_n^2 = N$, located at each site of the
spacetime lattice.

\subsection{Noninteracting  Theory}
\label{sec:noninteracting}
We first consider the limit $\lambda \to 0$ in Eq.~(\ref{Sorp}),
corresponding to the noninteracting field theory.  The rescaling factor
$K$ introduced in Eqs.~(\ref{K1})-(\ref{K3}) then becomes
\begin{equation} 
K = \frac{1}{2d + m^2 a^2}
\end{equation} 
and the action reduces to the quadratic form
\begin{equation} 
\label{Sl}
S_0 = \sum_{n, n'} \phi_n \cdot \Lambda_{n, n'} \phi_{n'} ,
\end{equation} 
with
\begin{equation}
\label{Lambda} 
\Lambda_{n, n'} =\delta_{n, n'} - K \sum_{\pm \mu} \delta_{n, n' +
\hat{\mu} }
\end{equation}
where the sum $\sum_{\pm \mu}$ extends over the positive as well as the
negative directions.  In matrix notation
\begin{equation} 
\Lambda =  I - K \, H,
\end{equation}
where $I$ is the identity matrix and $H$ denotes the so-called hopping
matrix\index{hopping!matrix} whose elements $H_{n, n'}$ are unity if the
two lattice sites $n$ and $n'$ are nearest neighbors, and zero
otherwise.  Physically, $H$ describes the hopping\index{hopping}, or
propagation of a particle from one lattice site, representing a
spacetime cell, to an adjacent one, with each hop carrying a weight $K$.
Since the theory is noninteracting, a particle is free to hop at
random to any of its nearest neighbors without restriction.  It can
and will in general revisit sites previously visited and can even hop
onto a site already occupied by another particle.  The particle
trajectories are therefore \textit{phantom} worldlines that can freely
intersect and share bonds of the lattice without energy penalty.

The inverse of $\Lambda$ determines the lattice correlation function:
\begin{equation} 
\label{lgreen}
\left\langle \phi^\alpha_n \phi^\beta_{n'} \right\rangle = 
\frac{1}{2} \delta_{\alpha, \beta}  \Lambda^{-1}_{n, n'} = 
\frac{1}{2} \delta_{\alpha, \beta}  \sum_{b=0}^{\infty} \, H^b_{n, n'}  K^b,
\end{equation} 
as can be read off from the action (\ref{Sl}).  From the definition of
the hopping matrix\index{hopping!matrix} $H$ it follows that $H_{n, n'}$
raised to the power $b$ gives the number of paths $z_b(n, n')$ of $b$
steps starting at site $n$ and ending at site $n'$.  In this so-called
\textit{hopping expansion}\index{hopping!expansion},\cite{GFCM} the
correlation function is therefore obtained by summing over all possible
worldlines of arbitrary many steps joining the endpoints $n$ and
$n'$:\cite{Ambjorn}
\begin{equation} 
\label{Sigma}
\left\langle \phi^\alpha_n \phi^\beta_{n'} \right\rangle = \frac{1}{2}
\delta_{\alpha, \beta} \sum_{b=0}^{\infty} z_b(n,n') K^b = \frac{1}{2}
\delta_{\alpha, \beta} \sum_\mathrm{worldlines} K^b .
\end{equation}
Each path carries a weight $K^b$ according to the number of steps $b$
taken.  This geometric representation of the correlation function as a
sum over worldlines constitutes Feynman's spacetime
approach\cite{Feynman48} to fluctuating fields on the lattice.  The
equivalence of the two approaches is known as \textit{particle-field
duality}\index{particle-field duality}.

Because of translational invariance, the number of particle trajectories
$z_b(n,n')$ depends only on the relative coordinate $z_b(n, n') = z_b(n-
n')$.  It satisfies the recurrence relation
\begin{equation}
\label{rec}
z_{b+1} (n) =  \sum_{\pm \mu} z_b (n+\hat{\mu}),
\end{equation} 
with the initial condition $z_0 (n) = \delta_{0,n}$.  This relation is
easily solved by going over to Fourier space
\begin{equation} 
z_b (n- n') =  \int_{-\pi}^{\pi} \frac{\mathrm{d}^d q}{(2 \pi)^d} \,
\mathrm{e}^{\mathrm{i} q \cdot (n- n')} z_b (q),
\end{equation} 
with $q \cdot n=q_\mu n_\mu = q_1 n_1 + \cdots q_d n_d$ and $q_\mu$ the
dimensionless momentum variable, yielding
\begin{equation} 
z_{b+1} (q) =  \sum_\mu \cos(q_\mu) z_b (q).
\end{equation}
With $z_0 (q) = 1$ as initial condition, this gives for $z_b(n-n')$
\begin{equation} 
\label{prob}
z_b (n-n') = \int_{-\pi}^{\pi} \frac{\mathrm{d}^d q}{(2
\pi)^d} \, \mathrm{e}^{\mathrm{i} q \cdot (n- n')} \left[
2\sum_\mu \cos (q_\mu) \right]^b.
\end{equation} 
In the limit where the lattice spacing $a$ is small but still finite, so
that the particles still take discrete steps, we have
\begin{eqnarray} 
\left[2K\sum_\mu \cos (q_\mu) \right]^b &\rightarrow&
\left(\frac{1 - q^2/2d}{1+m^2 a^2/2d} \right)^b
\nonumber \\ &\rightarrow& \exp \left[- (q^2 + m^2 a^2) b /2 d \right].
\end{eqnarray}  
The lattice correlation function (\ref{Sigma}) then becomes
\begin{eqnarray}
\label{discrete}  
\!\!\!\!\!\!\!\! \left\langle \phi^\alpha_n \phi^\beta_{n'}
\right\rangle &\rightarrow& \delta_{\alpha, \beta} \frac{1}{2}
\sum_{b=0}^{\infty} \int_{-\pi}^{\pi} \frac{\mathrm{d}^d q}{(2 \pi)^d}
\, \mathrm{e}^{\mathrm{i} q \cdot (n-n')} \, \mathrm{e}^{- (q^2 + m^2
a^2) b /2d} \nonumber \\ &\rightarrow& \delta_{\alpha, \beta}
\frac{1}{2} \sum_{b=1}^{\infty}  \left( \frac{d}{2 \pi b}
\right)^{d/2} \exp \left[- \frac{d}{2} \frac{(n - n')^2}{ b}\right] \,
\mathrm{e}^{- m^2 a^2 b/2d},
\end{eqnarray}
showing that the end-to-end vector $n-n'$ is Gaussian distributed with
an extra Boltzmann weight $\exp(-m^2 a^2/2d)$ associated with each step
taken.  Because of this bond fugacity, long worldlines are exponentially
suppressed as long as the mass term in the action is positive, $m^2>0$.
Notice that in the last line of Eq.~(\ref{discrete}), the lower bound on
the summation is replaced by $b=1$.  This is justified because the $b=0$
term always vanishes for $n \neq n'$.

A more familiar expression for the correlation function is obtained from
Eq.~(\ref{Sigma}) by first summing over the length $b$ of the particle
trajectories.  With $z_b(n,n')$ given in Eq.~(\ref{prob}), this leads to
\begin{eqnarray} 
\label{standard}
\left\langle \phi^\alpha_n \phi^\beta_{n'} \right\rangle &=&
  \delta_{\alpha, \beta} \frac{1}{2} \int_{-\pi}^{\pi}
  \frac{\mathrm{d}^d q}{(2 \pi)^d} \frac{\mathrm{e}^{\mathrm{i} q \cdot
  (n-n')}}{1-2 K \sum_\mu \cos (q_\mu)} \nonumber \\ &=&
  \delta_{\alpha, \beta} \frac{1}{2K} \int_{-\pi}^{\pi}
  \frac{\mathrm{d}^d q}{(2 \pi)^d} \frac{\mathrm{e}^{\mathrm{i} q \cdot
  (n-n')}}{2 \sum_\mu [1-\cos (q_\mu)] + m^2 a^2 }.
\end{eqnarray} 
For the original fields $\varphi_x$, which are connected to the rescaled
fields $\phi_n$ through Eq.~(\ref{K1}), this translates into the
standard form of the correlation function of a free field theory on a
hypercubic lattice,
\begin{equation} 
\label{hyper}
\left\langle \varphi^\alpha_x \varphi^\beta_{x'} \right\rangle = \delta_{\alpha,
\beta} a^2 \int_{-\pi/a}^{\pi/a} \frac{\mathrm{d}^d k}{(2 \pi)^d}
\frac{\mathrm{e}^{\mathrm{i} k \cdot (x-x')}}{2 \sum_\mu [1-\cos
(k_\mu a)] + m^2 a^2 },
\end{equation} 
where the lattice sites are now labeled again by $x_\mu= n_\mu a$, and
$k_\mu=q_\mu/a$ is the momentum variable.  

In the limit $a \to 0$, the correlation function (\ref{hyper}) reduces
to the standard continuum expression
\begin{equation} 
\left\langle \varphi^\alpha_x \varphi^\beta_{x'} \right\rangle = \delta_{\alpha,
\beta} \int_{-\infty}^{\infty} \frac{\mathrm{d}^d k}{(2 \pi)^d}
\frac{\mathrm{e}^{\mathrm{i} k \cdot (x-x')}}{k^2 + m^2}.
\end{equation} 
The momentum integral can be easily carried out by introducing the
Schwinger proper time representation\cite{proptime} of the integrand,
which is based on Euler's form
\begin{equation} 
\label{Euler}
\frac{1}{A^z} = \frac{1}{\Gamma(z)} \int_0^\infty \frac{\mathrm{d} s}{s}
\, s^z \, \mathrm{e}^{- s A},  
\end{equation} 
where $\Gamma(z)$ is the Euler gamma function.  Specifically,
\begin{eqnarray} 
\label{continuum}
\left\langle \varphi^\alpha_x \varphi^\beta_{x'} \right\rangle &=&
\delta_{\alpha, \beta} \int_0^{\infty} \mathrm{d} s \, \int
\frac{\mathrm{d}^d k}{(2 \pi)^d} \, \mathrm{e}^{\mathrm{i} k \cdot
(x-x')} \, \mathrm{e}^{-s \left(k^2+m^2\right)} \nonumber \\ &=&
\delta_{\alpha, \beta} \int_0^{\infty} \mathrm{d} s \, \left( \frac{1}{4
\pi s} \right)^{d/2}  \mathrm{e}^{-(x-x')^2/4s} \, \mathrm{e}^{-s m^2}
\nonumber \\ &=& \delta_{\alpha, \beta} \frac{1}{(2 \pi)^{d/2}}
\left(\frac{m}{|x-x'|}\right)^{d/2-1} K_{d/2-1}(m|x-x'|),
\end{eqnarray} 
with $K_{d/2-1}$ a modified Bessel function.  The integral in the second
line is the continuum limit of the lattice expression (\ref{discrete}),
with the sum over the number of steps $b$ taken by the particle replaced
by an integral over the Schwinger proper time parameter $s$, with the
identification
\begin{equation} 
\label{identification}
b a^2/2d \equiv s.
\end{equation}   
The parameter $s$ is related to the proper time $\tau_\mathrm{p}$ of the
trajectory traced out by the point particle, for which 
\begin{equation} 
\left(\frac{\mathrm{d} x_\mu(\tau_\mathrm{p})}{\mathrm{d}
\tau_\mathrm{p}} \right)^2=1.
\end{equation} 

To make this evident, we write the integrand in the second line of
Eq.~(\ref{continuum}) as a Feynman path integral\index{path
integral}\cite{StoneThomas}
\begin{multline}
\label{pathintegral}
\left( \frac{1}{4 \pi s} \right)^{d/2} \mathrm{e}^{-(x-x')^2/4s} \,
\mathrm{e}^{-s m^2} \\ = \int^{x(s) = x'}_{x(0) = x} \mathrm{D}x(s') \,
\exp \left\{-\int_0^{s}\mathrm{d} s' \left[ \tfrac{1}{4} {\dot
{x}}^2(s') + m^2 \right]\right\},
\end{multline} 
where  the so-called Wiener measure is defined by
\begin{equation} 
\int^{x(s) = x'}_{x(0)=x} \mathrm{D} x(s') \equiv  
\lim_{b\to \infty} 
A^b \prod_{i=1}^{b-1} \int_{-\infty}^\infty \mathrm{d}  x_i \, ,
\end{equation} 
with the normalization factor
\begin{equation} 
A = \left( \frac{d}{2 \pi a^2} \right)^{d/2}.
\end{equation} 
Here, it is understood that the limit $b \to \infty$ is taken
concurrently with the continuum limit $a \to 0$ in a way such that $b
a^2 \to \mbox{const}$.  The classical action of the point particle in
Eq.~(\ref{pathintegral}), tracing out its trajectory in spacetime is the
continuum limit of the lattice action present in Eq.~(\ref{discrete}):
\begin{eqnarray} 
\label{W0}
W_0 &=& \frac{d}{2} \sum_{i=1}^b \left(\frac{x_{i-1} - x_i}{a}\right)^2 +
\frac{m^2 a^2 b}{2d} \nonumber \\ &=& \sum_{i=1}^b \frac{a^2}{2d}
\left[\frac{1}{4} \left(\frac{x_{i-1} - x_i}{a^2/2d}\right)^2 + m^2
\right] \nonumber \\ &\to& \int_0^{s} \mathrm{d} s' \left[\tfrac{1}{4}
\dot{x}^2(s') + m^2 \right],
\end{eqnarray} 
where in the last step the identification (\ref{identification}) is
used.  Written as a path integral\index{path integral}, the correlation
function (\ref{continuum}) thus reads
\begin{equation} 
\left\langle \varphi^\alpha_x \varphi^\beta_{x'} \right\rangle = \delta_{\alpha,
\beta} \int_0^{\infty} \mathrm{d} s \, 
\int^{x(s) = x'}_{x(0) = x} \mathrm{D}x(s') \, \mathrm{e}^{- W_0} .
\end{equation} 
It forms the continuum limit of the sum over all possible particle
trajectories on the spacetime lattice in Eq.~(\ref{Sigma}).  Note that
the proper time parameter $s$ can take any positive value, in agreement
with the lattice representation (\ref{Sigma}) of the correlation
function where the worldlines can be arbitrarily long.  The Boltzmann
factor $\exp (-s m^2)$ exponentially suppresses, however, large proper
times as long as $m^2>0$.  When we set
\begin{equation} 
s \equiv \tau_\mathrm{p}/2m ,
\end{equation} 
the continuum action (\ref{W0}) yields $m$ times the arc~length of the
path,
\begin{equation} 
\int_0^{s} \mathrm{d} s' \left[\tfrac{1}{4}
\dot{x}^2(s') + m^2 \right] = m \tau_\mathrm{p},
\end{equation} 
which is the standard form of the classical action of a relativistic
free scalar particle of mass $m$.  This demonstrates that $s$ can indeed
be thought of as the proper time.

For later use, we record the second line in Eq.~(\ref{continuum}) with
Planck's constant and the speed of light reinstalled:
\begin{equation} 
\label{chbar}
\left\langle \varphi^\alpha_x \varphi^\beta_{0} \right\rangle =
\delta_{\alpha, \beta} \, c \int_0^{\infty} \mathrm{d} s \, \left(
\frac{1}{4 \pi \hbar s} \right)^{d/2} \mathrm{e}^{-(c^2 \tau^2 +
\mathbf{x}^2)/4 \hbar s} \, \mathrm{e}^{-s m^2 c^2/\hbar} ,
\end{equation} 
while the continuum action (\ref{Scont}) becomes
\begin{equation} 
\label{relS}
S_0 = \int \mathrm{d}  \tau \, \mathrm{d}^{d-1} x \; \varphi \left( -
\hbar^2 \frac{\partial^2}{c^2\partial \tau^2} - \hbar^2\nabla^2 + m^2 c^2
\right)\varphi,
\end{equation} 
where now the spacetime coordinates are denoted by $x_\mu = ({\bf x},c
\tau)$, with $\tau$ the imaginary time and $\mathbf{x}=(x_1,
\cdots,x_{d-1})$ the spatial coordinates.

Besides the number of worldlines $z_b(n, n')$ on the spacetime lattice
connecting the lattice sites $n$ and $n'$ in $b$ steps, also the
probability $P_b(n, n')$ for a particle to move from $n$ to $n'$ in
$b$ steps will become important below. The probability is given by the
ratio
\begin{equation} 
\label{Pn}
P_b(n, n') = z_b(n,n')/z_b
\end{equation} 
of $z_b(n, n')$ and the number of paths of $b$ steps starting at $n$ and
ending at an arbitrary lattice site,
\begin{equation} 
\label{zb}
z_b \equiv \sum_{n'} z_b(n,n').
\end{equation}   
The denominator in Eq.~(\ref{Pn}) is included so that $P_b$ is a genuine
probability, taking values between 0 and 1.  To evaluate it, imagine the
particle taking a first step to one of its nearest neighbors $n
\pm\hat{\mu}$.  It then has only $n-$1 steps still available to reach
the final destination $n'$ with probability $P_{b-1}(n \pm \hat{\mu},
n')$.  That is, $P_b$ satisfies the recurrence relation:
\begin{equation} 
\label{recu}
P_b(n, n') = \frac{1}{2d} \sum_{\pm \mu}  P_{b-1}(n+ \hat{\mu}, n')
\end{equation} 
where the factor $2d$ denotes the number of nearest neighbors on a
$d$-dimensional square lattice.  When $P_{b-1}(n, n')$ is
subtracted from both sides of Eq.~(\ref{recu}), the difference equation
\begin{equation} 
P_b(n, n') - P_{b-1}(n, n') = \frac{1}{2d} \sum_{\pm \mu}
\left[P_{b-1}(n + \hat{\mu}, n') - P_{b-1}(n, n') \right]
\end{equation} 
follows, with the boundary condition $P_0 (n, n') = \delta_{n,n'}$.  As
before, the continuum limit is taken by simultaneously letting $a \to 0$
and $b \to \infty$, such that $b a^2 \to \mbox{const}$.  The difference
equation then turns into the differential equation
\begin{equation} 
\label{diff}
\partial_b P_b(n, n') = \frac{a^2}{2d} \nabla^2 P_b(n, n'),
\end{equation} 
with the celebrated solution 
\begin{equation} 
\label{celebrate}
P_b(n,n') = \left( \frac{d}{2 \pi b} \right)^{d/2} \exp \left[-
\frac{d}{2} \frac{(n - n')^2}{b}\right].
\end{equation} 
For the probability density $p_b(x, x') \equiv P_b(n,n')/a^d$ this gives
the Gaussian distribution
\begin{equation} 
\label{celebrated}
p_b(x,x') = \left( \frac{d}{2 \pi b a^2} \right)^{d/2} \exp \left[-
\frac{d}{2} \frac{(x - x')^2}{b a^2}\right].
\end{equation} 
Each dimension the particle is free to roam contributes a factor
$\sqrt{d/2 \pi b a^2}$ to the prefactor.

\subsection{Loop Gas}
\label{sec:loopgas}
\begin{figure}
\centering
\psfrag{x}[t][t][.8][0]{$n$}
\includegraphics[width=0.5\textwidth]{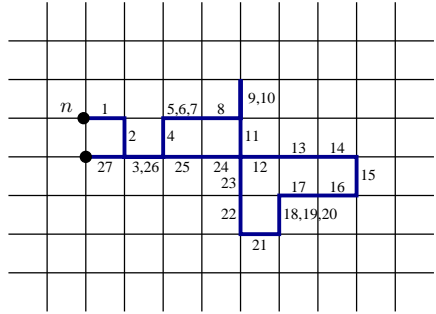}
\caption{A Brownian random walk on a square lattice returning to a site
adjacent to its starting position $n$.  As indicated, the particle took 27
steps in total.  Observe that various bonds are traversed more than
once.
  \label{fig:rw}}
\end{figure}

A similar geometric representation in terms of particle trajectories as
found for the correlation function can be given for the partition
function.  Consider the partition function of the free
theory:
\begin{equation} 
  Z_0 = \prod_n \int \mathrm{d} \phi_n \; \mathrm{e}^{-S_0},
\end{equation} 
with the lattice action $S_0$ given by Eq.~(\ref{Sl}).  
This integral generalizes the standard Gaussian integral
\begin{equation} 
\label{gaussR}
\int \prod_{i=1}^n \mathrm{d} x_i \exp \left(-\frac{1}{2} \sum_{i,j=1}^n
x_i M_{i j} x_j \right) = (2 \pi)^{n/2} \, {\rm det}^{-1/2} (M),
\end{equation} 
with $M$ a symmetric positive-definite $n\times n$ matrix, and ${\rm
det} (M)$ its determinant. It can therefore be evaluated exactly, with
the result 
\begin{equation} 
\label{Z00}
\ln Z_0 = \ln \mathrm{Det}^{-N/2} (\Lambda) = - \frac{N}{2} \mathrm{Tr} \ln
\Lambda ,
\end{equation} 
where Tr denotes the functional trace. In deriving this, irrelevant
prefactors are ignored and use is made of the identity 
\begin{equation} 
\ln \mathrm{Det}(A) = \mathrm{Tr} \ln A, 
\end{equation} 
relating the determinant and trace of an Hermitian operator $A$.  The
relation can be easily checked explicitly for a $n\times n$ matrix.
With $\Lambda = I - K H$ as before, $\ln Z_0$ can be expanded in the
hopping parameter\index{hopping!parameter} $K$ as
\begin{equation} 
\label{Z0}
\ln Z_0 = \frac{N}{2} L^d \sum_{b=1}^{\infty} \frac{1}{b} z_b(a)  K^b
\end{equation} 
where $z_b(a)$ denotes the number of worldlines of $b$ steps, starting
at a given lattice site $n$ and returning to one of its nearest neighbor
sites $n \pm \hat{\mu}$ after $b$ steps (see Fig.~\ref{fig:rw}).  The
lattice spacing $a$ serves here as a microscopic cutoff, so that
$z_b(a)$ rather than $z_b(0)$ appears when closing open paths.  Since a
worldline can start at any site on the spacetime lattice, the right hand
contains a factor $L^d$, denoting the total number of lattice sites, and
$\ln Z_0$ is extensive.
The factor $1/2b$ appearing in Eq.~(\ref{Z0}) reflects the possibility
to trace out a closed path starting from any site along the path and
going in either direction round it.  The factor $N$ reflects the
degeneracy associated with a closed worldline.  Written as a sum over
\textit{single} loops of arbitrary length and shape, Eq.~(\ref{Z0}) reads
\begin{equation} 
\label{lnZcomp}
\ln Z_0 = N \sum_{\substack{\mathrm{single} \\ \mathrm{loops}}} K^b,
\end{equation} 
where the loops are considered to have no orientation, whence the
absence of the factor $\frac{1}{2}$, and no longer rooted, whence the
absence of the factor $L^d$.

For a noninteracting theory, $z_b(a)$ can be obtained simply from
Eq.~(\ref{prob}) by putting the end-to-end vector $n-n'$ to zero there.
Equation (\ref{Z0}) then reads explicitly
\begin{equation} 
\label{lnZ0s}
  \ln Z_0 = \frac{N}{2} L^d \int_{-\pi}^{\pi} \frac{\mathrm{d}^d q}{(2
  \pi)^d} \sum_{b=1}^{\infty} \frac{1}{b} \left[ 2 K\sum_\mu \cos
  (q_\mu) \right]^b.
\end{equation} 
Repeating the same steps leading to Eq.~(\ref{discrete}), one finds
\begin{eqnarray} 
\label{lnZ0}
  \ln Z_0 &=& \frac{N}{2} L^d \sum_{b=1}^{\infty} \int_{-\pi}^{\pi}
  \frac{\mathrm{d}^d q}{(2 \pi)^d} \frac{1}{b} \, \mathrm{e}^{- (q^2 +
  m^2 a^2) b /2d} \nonumber \\ &=& \frac{N}{2} L^d \sum_{b=1}^{\infty}
  \frac{1}{b} \, \left( \frac{d}{2 \pi b} \right)^{d/2} \mathrm{e}^{-
  m^2 a^2 b/2d} .
\end{eqnarray}
It follows that the logarithm of the partition function has the
form\cite{Copeland:1988bh,Copetal}
\begin{equation} 
\label{Zell}
\ln Z_0/L^d \sim \sum_b \ell_b,
\end{equation} 
with the loop distribution\index{loop!distribution}
\begin{equation}
\label{ell}
\ell_b \sim b^{- \tau} \, \mathrm{e}^{- \theta b}.
\end{equation}
denoting the average number (per lattice site) of times a closed
worldline of $b$ steps occurs on the lattice.  The parameter $\theta
\propto m^2$ physically determines the line tension of the loops, while
the algebraic factor in the loop distribution\index{loop!distribution},
characterized by the exponent $\tau=d/2+1$, is an entropy factor, giving
a measure of the number of ways a loop of $b$ steps can be embedded in
the lattice.

A more familiar expression for $\ln Z_0$ is obtained by again first
carrying out the sum over the loop length $b$ in Eq.~(\ref{lnZ0s}):
\begin{eqnarray} 
 \ln Z_0 &=& - \frac{N}{2} L^d \int_{-\pi}^{\pi} \frac{\mathrm{d}^d
  q}{(2 \pi)^d} \ln \left[1-2 K \sum_\mu \cos (q_\mu) \right] \nonumber
  \\ &=& - \frac{N}{2} \Omega \int_{-\pi/a}^{\pi/a} \frac{\mathrm{d}^d k}{(2
  \pi)^d} \ln \left\{2 \sum_\mu \left[1-\cos \left(k_\mu a\right)\right]
  + m^2 a^2 \right\},
\end{eqnarray} 
where in the last step an irrelevant additive factor is ignored and
$\Omega=(L a)^d$ is the volume of the spacetime box.  This is the standard
representation of the logarithm of the partition function describing a
free field theory on a hypercubic lattice.  In the continuum limit $a
\to 0$, the momentum integral can again best be carried out by using the
Schwinger proper time representation this time of the logarithm,
\begin{equation} 
\label{lnS}
\ln(a) = \lim_{z \to 0} \frac{1}{z} - \int_0^\infty
\frac{\mathrm{d}s}{s} \, \mathrm{e}^{-s a},
\end{equation} 
where the first term is an irrelevant diverging contribution.  Apart
from further irrelevant additive constants this leads to 
\begin{eqnarray}
\label{Z_0S} 
\ln Z_0 &=& \ln \mathrm{Det}^{-N/2}\left( - \partial_\mu^2 + m^2 \right)
  \nonumber \\ &=& - \frac{N}{2} \Omega \int_{-\infty}^\infty
  \frac{\mathrm{d}^d k}{(2 \pi)^d} \ln \left(k^2 + m^2 \right) \nonumber
  \\ &=& \frac{N}{2} \Omega \int_0^\infty \frac{\mathrm{d}s}{s} \,
  \left(\frac{1}{4 \pi s}\right)^{d/2} \mathrm{e}^{-s m^2} ,
\end{eqnarray}  
where the operator $- \partial_\mu^2 + m^2$ is the continuum limit of
$\Lambda$ introduced in Eq.~(\ref{Lambda}).  The right hand of
Eq.~(\ref{Z_0S}) forms the continuum limit of the lattice expression in
the last line of Eq.~(\ref{lnZ0}) with the number of steps $b$ taken by
the particle assuming the role of the Schwinger proper time parameter
$s$: $b a^2/2d \equiv s$ as in Eq.~(\ref{identification}).

On account of the path integral\index{path integral}
(\ref{pathintegral}) with $x=x'$, the last two factors in the integrand
at the right hand of Eq.~(\ref{Z_0S}) can be written as a path
integral\index{path integral} over \textit{closed} worldlines
\begin{equation} 
\label{pathintegralo}
\mathrm{e}^{-s m^2} \left( \frac{1}{4 \pi s} \right)^{d/2} = \oint
\mathrm{D}x(s') \, \exp\left\{-\int_0^{s}\mathrm{d} s'
\left[\tfrac{1}{4} {\dot {x}}^2(s') + m^2 \right] \right\},
\end{equation} 
where the notation $\oint \mathrm{D} x(s')$ refers to closed paths.  In
this way, $\ln Z_0$ expressed as a path integral\index{path integral}
reads\cite{StoneThomas}
\begin{eqnarray}  
\label{lnZ0pi}
\ln Z_0 &=& \ln \mathrm{Det}^{-N/2}\left( - \partial_\mu^2 + m^2 \right)
   \nonumber \\ &=& \frac{N}{2}  \Omega \int_0^\infty \frac{\mathrm{d}s}{s}
   \oint \mathrm{D}x(s') \, \exp\left\{-\int_0^{s} \mathrm{d} s'
   \left[\tfrac{1}{4} \dot{x}^2(s') + m^2 \right] \right\},
\end{eqnarray} 
where the argument of the exponential function is given by (minus) the
classical action (\ref{W0}) of the point particle tracing out its
trajectories in spacetime.  An easy mnemonic for remembering the main
ingredients of this formula is to interpret $- \partial_\mu^2 + m^2$ as
the Hamilton operator describing the motion of a
\textit{nonrelativistic} particle of mass $M=\tfrac{1}{2}$ in a constant
potential $V=m^2$ in $d$ dimensions.  With the Schwinger proper time $s$
interpreted as Newtonian time, $W_0$ is the corresponding
nonrelativistic action, where it is recalled that in natural units
$\hbar =1$.

The first line in Eq.~(\ref{Z_0S}) is sometimes represented by the
one-loop Feynman diagram
\begin{equation} 
\ln Z_0 = \bigcirc.
\end{equation}   
In the spacetime approach, this diagram stands for all the terms in the
sum (\ref{Z0}), each consisting of a \textit{single} worldline starting
and ending at a given site $n$.  That is, the Feynman diagram at the
same time denotes the topology of the particle trajectories contributing
to $\ln Z_0$.\cite{GFCM}

The partition function, obtained by exponentiating Eq.~(\ref{Z0}), can
be written in terms of Feynman diagrams as
\begin{equation} 
\label{loopgas}
  Z_0 = 1 + \bigcirc + \frac{1}{2!} \bigcirc \bigcirc + \frac{1}{3!} \bigcirc 
  \bigcirc \bigcirc + \cdots 
\end{equation} 
Each diagram involves a separate integration over loop momentum $k_\mu$.
With the right hand of Eq.~(\ref{loopgas}) taken as picturing the
topology of the particle trajectories contributing to the partition
function, it follows that $Z_0$ describes a loop gas\index{loop!gas}--a
grand canonical ensemble of fluctuation loops of arbitrary shape and
length.\cite{Samuel} Being a noninteracting theory, the closed
worldlines can freely intersect themselves or other worldlines and even
share bonds.  On the spacetime lattice, the loop gas\index{loop!gas}
(\ref{loopgas}) can be expressed as a sum over all possible tangles of
closed paths
\begin{equation} 
\label{phantom} 
Z_0 = \sum_\mathrm{loops} K^b N^l,
\end{equation} 
with $b$ and $l$ denoting the total number of occupied bonds and
separate loops forming a given tangle. In Eq.~(\ref{phantom}), the loops
are no longer considered to be rooted, i.e., they can start and end at
any spacetime coordinate in the system.  In comparison with rooted
closed paths, i.e., closed paths all starting at a given spacetime
coordinate, this yields a spacetime volume factor $\Omega$.  In the
continuum, Eq.~(\ref{phantom}) translates into the path
integral\index{path integral} representation
\begin{equation} 
\label{explicit}
Z_0 = \sum_{l=0}^{\infty} \frac{1}{l!}  \left(\frac{N}{2}\right)^l
\prod_{r=1}^l \left[ \Omega \int_0^\infty \frac{\mathrm{d} s_r}{s_r}  \oint
\mathrm{D} x_r(s'_r) \right] \mathrm{e}^{-W^{(l)}_0},
\end{equation} 
as follows from Eq.~(\ref{lnZ0pi}).  Here, $W^{(l)}_0$ denotes the extension
of the single particle action (\ref{W0}) to $l$ particles:
\begin{equation} 
\label{Wl0}
W^{(l)}_0 = \sum_{r=1}^l \int_0^{s_r} \mathrm{d} s'_r \,
\left[\tfrac{1}{4} \dot{x}_r^2 (s'_r) + m^2 \right].
\end{equation} 

\subsection{$|\phi|^4$ Field Theory\index{phi@$\phi^4$ theory}}
\label{sec:phi4}
\begin{figure}
\begin{center}
\includegraphics[width=.8\textwidth]{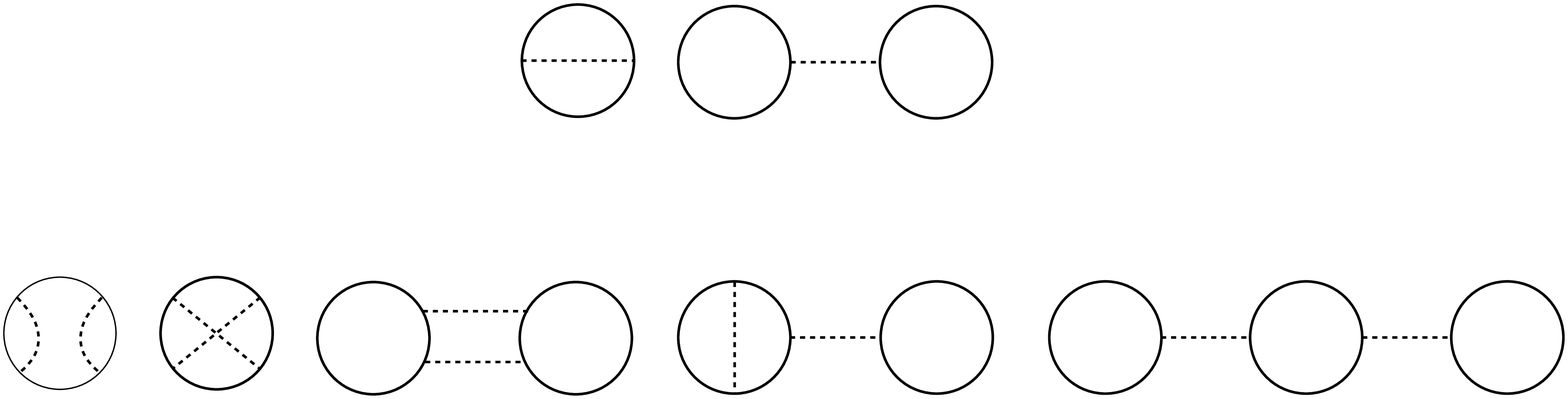}
\end{center}
\caption{Two- and 3-loop Feynman diagrams contributing to $\ln Z$ of
the $|\varphi|^4$ theory.
  \label{fig:23loop}}
\end{figure}

We next perturbatively include the interaction in the theory
(\ref{Sor}).  At the level of the partition function (\ref{Z}), this
boils down to expanding the interaction term of the action in a Taylor
series
\begin{equation} 
\label{Taylor}
\mathrm{e}^{- (g/4!) a^d \sum_x  \varphi^4_x} = 1 - \frac{g}{4!} 
a^d \sum_x  \varphi^4_x 
+ \frac{1}{2!} \left(\frac{g}{4!}\right)^2 a^{2d} \sum_{x,x'}  
\varphi^4_x \varphi^4_{x'} + \cdots .
\end{equation} 
The terms in this expansion can all be represented by Feynman
diagrams.\cite{Amit,KlSF} Figure~\ref{fig:23loop} shows the 2- and
3-loop diagrams contributing to $\ln Z$ as an example.  Each loop, which
are considered to have no orientation, carries a factor of $N$.  The
contact interaction is represented by a dashed line for convenience,
each carrying a weight $\propto g$.  In the equivalent spacetime
approach, Feynman diagrams are understood as spacetime diagrams.  Each
loop of a Feynman diagram is expressed in a hopping
expansion\index{hopping!expansion} as in the noninteracting theory, with
the interaction standing for an intersection between different
worldlines or between different parts of a single worldline
(self-intersection).  In addition to the ones indicated, a worldline
contributing to a given Feynman diagram may have further intersections,
but these are treated as in the noninteracting case, i.e., as
nonexisting.  Put differently, apart from the intersections explicitly
indicated, the particle trajectories still behave as phantom paths.

Picturing the topology of closed worldlines, the first Feynman diagram
($\propto N$) of the 2-loop contributions ($\propto g^2$) in
Fig.~\ref{fig:23loop} represents single closed particle trajectories of
arbitrary length and shape with one self-intersection carrying a factor
of $g$.  The second Feynman diagram ($\propto N^2$) represents two
intersecting closed worldlines.  The 3-loop contributions ($\propto
g^3$) represent single closed worldlines with two self-intersections
(1st and 2nd diagram $\propto N$), two closed worldlines intersecting
twice (3rd diagram $\propto N^2$), two closed worldlines intersecting
once with one of the two having in addition one self-intersection (4th
diagram $\propto N^2$), and finally three closed worldlines (last
diagram $\propto N^3$) connected by two intersections.

Similarly, the $n$th order contributions to $\ln Z$, corresponding to
the ($n-1$)th term ($\propto g^{n-1}$) in the Taylor expansion
(\ref{Taylor}), all involve a total of $n-1$ intersections (including
self-intersections) and up to $n$ separate loops.  The resulting loop
gas is similar to that of the free theory, but the random loops are no
longer phantom loops as the intersections explicitly indicated carry a
weight $\propto g$.  In the path integral\index{path integral}
formulation, this is reflected by an additional steric repulsion in the
worldline action,
\begin{eqnarray} 
\label{steric}
W^{(l)} &=& \sum_{r=1}^l \int_0^{s_r} \mathrm{d} s'_r \left[\tfrac{1}{4}
\dot{x}_r^2(s'_r) + m^2 \right]  \nonumber \\ && + \frac{g}{6}
\sum_{r,r'=1}^l \int_0^{s_r} \mathrm{d} s'_r \int_0^{s_{r'}} \mathrm{d}
s'_{r'} \, \delta \left[x_r(s'_r) - x_{r'} (s'_{r'}) \right],
\end{eqnarray} 
where the sums $\sum_r$ and $\sum_{r'}$ extend over the $l$ particle
trajectories assumed to be present. The net result is that
(self-)intersections are suppressed and the random loops tend to become
mutually and self-avoiding, the more so, the larger the coupling constant
$g$ becomes.  Turning on the interaction thus changes the typical form
of the random loops and in particular their fractal structure.

To derive Eq.~(\ref{steric}), we follow Symanzik\cite{Symanzik} and
write the interaction term in the action (\ref{Sor}) in the continuum
limit as a functional integral\index{functional integral} over an
auxiliary field $\sigma$
\begin{equation} 
\exp \left( - \frac{g}{4!} \int \mathrm{d}^d x \, \varphi^4 \right) =
\int \mathrm{D} \sigma \exp \left[ -\int \mathrm{d}^d x \left(
\frac{6}{g} \sigma^2 - \mathrm{i} \sigma \varphi^2 \right) \right],
\end{equation} 
up to an irrelevant prefactor.  The partition function (\ref{Z}) then
becomes in the continuum limit
\begin{eqnarray} 
Z &=& \int \mathrm{D} \varphi \, \mathrm{D} \sigma \exp \left\{- \int
\mathrm{d}^d x \left[ \frac{1}{2} (\partial_\mu \varphi)^2 +
\frac{m^2}{2} \varphi^2 - \mathrm{i} \sigma \varphi^2 + \frac{6}{g}
\sigma^2 \right] \right\} \nonumber \\ &=& \int \mathrm{D} \sigma \,
{\rm Det}^{-N/2} \left(- \partial_\mu^2 + m^2 - 2 \mathrm{i} \sigma
\right) \, \exp\left(-\frac{6}{g} \int \mathrm{d}^d x \, \sigma^2\right)
\nonumber \\ &=& \int \mathrm{D} \sigma \sum_{l=0}^{\infty} \frac{1}{l!}
\left(\frac{N}{2} \right)^l \prod_{r=1}^l \left[ \Omega \int_0^\infty
\frac{\mathrm{d} s_r}{s_r}  \oint \mathrm{D} x_r(s'_r) \right]
\exp\left(-\frac{6}{g} \int \mathrm{d}^d x \, \sigma^2\right) \nonumber
\\ && \quad \quad \quad \times \exp \left( - \sum_{r=1}^l \int_0^{s_r}
\mathrm{d} s'_r \left\{ \tfrac{1}{4} \dot{x}_r^2(s'_r) + m^2 - 2
\mathrm{i} \sigma \left[x_r (s'_r)\right] \right\} \right)
\end{eqnarray} 
where the last equality follows from the previous result
(\ref{explicit}).  The auxiliary field $\sigma$ is a function of the
coordinates along the particle trajectories, $\sigma = \sigma\left[x_r
(s'_r)\right]$.  A simple Gaussian integration yields
\begin{multline} 
\int \mathrm{D} \sigma \exp \left\{ -\frac{6}{g} \int \mathrm{d}^d x \,
\sigma^2 + 2 \mathrm{i} \sum_{r=1}^l \int_0^{s_r} \mathrm{d} s_r' \,
\sigma \left[ x_r(s_r') \right] \right\} \\ = \exp \left\{ - \frac{g}{6}
\sum_{r,r'=1}^l \int_0^{s_r} \mathrm{d} s_r' \int_0^{s_{r'}} \mathrm{d}
s_{r'}' \, \delta \left[ x (s_r') - x (s_{r'}') \right] \right\},
\end{multline} 
again up to an irrelevant prefactor.  Using this in the last expression
for $Z$, we obtain the spacetime representation of the partition
function of the $\varphi^4$ theory\index{phi@$\phi^4$ theory}
\begin{equation} 
\label{loopgasp}
Z = \sum_{l=0}^{\infty} \frac{1}{l!} \left(\frac{N}{2} \right)^l
\prod_{r=1}^l \left[ \Omega \int_0^\infty \frac{\mathrm{d} s_r}{s_r}  \oint
\mathrm{D} x_r(s'_r) \right] \, \mathrm{e}^{-W^{(l)}},
\end{equation} 
with the classical action $W^{(l)}$ of the point particles tracing out
their worldlines given in Eq.~(\ref{steric}).  The $l$th term in this
sum describes a tangle of $l$ loops.  Since the number of loops can be
arbitrary large, Eq.~(\ref{loopgasp}) denotes the partition function of a
grand canonical ensemble of fluctuation loops, each of arbitrary
length and shape, with steric repulsion.

\begin{figure}
\begin{center}
\psfrag{i}[t][t][.8][0]{$\alpha$}
\psfrag{j}[t][t][.8][0]{$\beta$}
\psfrag{+}[t][t][.8][0]{+}
\psfrag{1}[t][t][.8][0]{1}
\psfrag{2}[t][t][.8][0]{2}
\psfrag{3}[t][t][.8][0]{3}
\psfrag{4}[t][t][.8][0]{4}
\includegraphics[width=.6\textwidth]{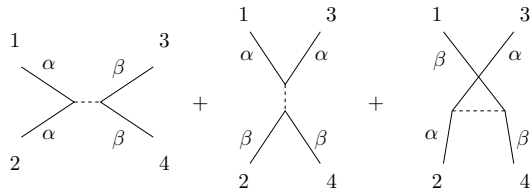}
\end{center}
\caption{The representation of the vertex of the $\varphi^4$
  theory\index{phi@$\phi^4$ theory} by a dashed line makes the routings
  of the field components, labeled by $\alpha$ and $\beta$, explicit.
  \label{fig:vertex}}
\end{figure}

Surprisingly, as was first pointed out by Balian and
Toulouse,\cite{BalianToulouse} the noninteracting theory can be
recovered from the interacting theory by taking the number of field
components $N \to -2$.  Consider a contribution to the partition
function involving the first vertex in Fig.~\ref{fig:vertex}.  Assume
that leg 1 is connected through the rest of the Feynman diagram to leg
2. Both legs carry the same spin index $\alpha$, which after taking the
sum yields a factor of $N$.  Legs 3 and 4 are necessarily also connected
through the rest of the Feynman diagram\index{Feynman diagrams|)}.  They
carry the same spin index $\beta$, say, which after taking the sum also
yields a factor of $N$.  Replacing the first vertex with the second or
third vertex in Fig.~\ref{fig:vertex} yields identical contributions
save for a change in topology regarding the routings of the spin
indices.  Specifically, the graph with the first vertex involves two
different sets of spin indices: $\alpha$ and $\beta$, whereas the graphs
with the second or third vertex involve only a single set of spin
indices as $\alpha=\beta$.  Since the first graph carries an extra
factor of $N=-2$ (the loop fugacity) after the sum over the spin index
is taken, the three graphs cancel.  All other graphs involving vertices
cancel in the same fashion three by three.  What remains are phantom
loops of a noninteracting theory.

\subsection{\mbox{O($N$)} Spin Model}
\label{sec:spinmodel}
Whereas the free theory discussed in Sec.~\ref{sec:noninteracting}
corresponds to taking the coupling constant $\lambda \to 0$ in the
action (\ref{Sorp}), the O($N$) spin model\index{O($N$) model}
corresponds to taking the opposite limit, $\lambda \to \infty$.  As for
$\lambda \to 0$, the theory becomes quadratic also in this limit of
infinite repulsion.  The theory is nevertheless nontrivial because of
the fixed-length constraint $\phi_n^2 = N$ at each lattice site.
Instead of considering the conventional Boltzmann weight $\exp(K\phi_n
\cdot \phi_{n'})$, often a simplified representative of the O($N$)
universality class is studied, obtained by truncating that
weight\cite{Nienhuis}
\begin{equation}
  \label{ZS}
  Z = \mathrm{Tr} \prod_{\langle n,n' \rangle} (1 + K \phi_n \cdot
  \phi_{n'}) .
\end{equation} 
The product is restricted to nearest neighbor pairs.  The remaining
factor in the on-site measure (\ref{onsite}) becomes trivial in this
limit and has been ignored.  The main difference with the original spin
model is that in the truncated model bonds cannot be multiply occupied.
It is generally accepted that this simplification does not change the
universal properties of the theory.

\begin{figure}
\begin{center}
\includegraphics[width=.4\textwidth]{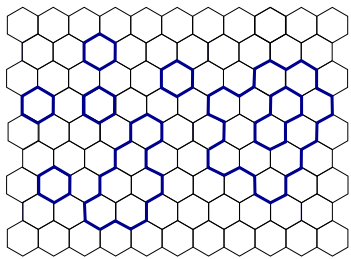}
\end{center}
\caption{Closed high-temperature graphs on a honeycomb lattice
automatically form mutually and self-avoiding loops.
  \label{fig:honeycomb}}
\end{figure}

The hopping expansion\index{hopping!expansion} is usually referred to
the \textit{high-temperature expansion} in the context of spin models.
Also for the truncated model, it corresponds to expanding the partition
function in terms of the parameter $K$.  The contributions to $Z$ can
again be visualized by graphs on the lattice.\cite{Stanley} As will be
detailed below for the Ising model\index{Ising model}, only closed
graphs with an even number of occupied bonds connecting each vertex
yield nonzero contributions.  The virtue of the truncated model is that
bonds cannot be multiply occupied.  In contrast, no restriction on
multiple occupancies exists in the representation with the usual
Boltzmann weight.  When formulated on a two-dimensional honeycomb
lattice, which has coordination number $z=3$, the high-temperature
graphs of the truncated model are automatically mutually and
self-avoiding:\cite{Nienhuis} a given lattice site is either empty or
has one bond entering and one leaving the site (see
Fig.~\ref{fig:honeycomb}).

The partition function can then be written geometrically as a sum over
all possible mutually and self-avoiding (MSA) loops,\cite{Nienhuis}
\begin{equation} 
\label{Zhoneycomb}
Z = \sum_{\substack{\mathrm{MSA} \\ \mathrm{loops}}} N^l K^b,
\end{equation}  
with $b$ and $l$ the number of occupied bonds and separate loops forming
the graph.  Equation (\ref{Zhoneycomb}) constitutes the high-temperature
representation of the partition function.  As for the noninteracting
field theory (\ref{phantom}), each occupied bond carries a weight $K$
(bond fugacity) and each loop carries a weight $N$ (loop fugacity).
Whereas the random loops of the noninteracting field theory are phantom
loops that can freely intersect and share bonds, here they are mutually
and self-avoiding.  As before, they physically denote the worldlines of
the particles described by the quantum field theory.

The O($N$) spin-spin correlation function 
\begin{equation} 
\left\langle \phi^\alpha_n \phi^\beta_{n'} \right\rangle = \frac{1}{Z}
 \prod_m \int \mathrm{d} \phi_m \; \phi^\alpha_n \phi^\beta_{n'} \, \exp
 \left(K \sum_{\langle m,m' \rangle} \phi_m \cdot \phi_{m'} \right),
\end{equation} 
is represented diagrammatically by a modified partition function,
obtained by requiring that the two sites $n$ and $n'$
are connected by an open high-temperature graph.\cite{Stanley}  On a
honeycomb lattice, the scaling part of the correlation function is given
by the connected graphs
\begin{equation}
\label{GR}
\left\langle \phi^\alpha_n \phi^\beta_{n'} \right\rangle \sim
  \frac{1}{2} \delta_{\alpha,\beta} \sum_\mathrm{graphs} K^b =
  \frac{1}{2} \delta_{\alpha,\beta} \sum_b z_b(n,n') K^b,
\end{equation} 
where $z_b(n,n')$ is the number of (open) mutually and self-avoiding
graphs along $b$ bonds connecting the lattice sites $n$ and $n'$.
Strictly speaking, $\left\langle \phi^\alpha_n \phi^\beta_{n'}
\right\rangle < \frac{1}{2} \delta_{\alpha,\beta} \sum_\mathrm{graphs}
K^b$ as the cancellation of the disconnected graphs in the numerator and
$Z$ in the denominator, required for an equality, is not complete: For a
given open graph, certain loops present in $Z$ are forbidden in the
modified partition function as they would intersect the open graph, or
occupy bonds belonging to it, which is not allowed.  In other words, the
presence of an open graph influences the loop gas\index{loop!gas} and
\textit{vice versa}.  Since each loop carries a factor $N$, the loop
gas\index{loop!gas} is absent in the limit $N \to 0$, pioneered by de
Gennes.\cite{deGennes} The inequality in Eq.~(\ref{GR}) becomes an
equality in this limit, and the open graphs become ordinary
self-avoiding random walks on a honeycomb lattice.  These walks
physically describe polymers in good solvents at sufficiently high
temperatures so that the van der Waals attraction between monomers is
irrelevant.  For the noninteracting theory, where the worldlines are
phantom trajectories, the loops cancel, so that the corresponding
correlation function is exactly given by the open trajectories
connecting the endpoints [see Eq.~(\ref{Sigma})].

For a $d$-dimensional hypercubic lattice, the coordination number is
$z=2d$, so that intersections are now possible.  Although the bonds of
the truncated model (\ref{ZS}) can by construction still not be multiply
occupied, the high-temperature graphs are no longer mutually and
self-avoiding.  After rescaling the bond fugacity $K \to K/N$, the
partition function (\ref{ZS}) can be written as\cite{CPS}
\begin{equation} 
\label{Zsquare}
Z = \sum_\mathrm{loops} \nu_2^{m_2} \nu_4^{m_4} \cdots \nu_{2d}^{m_{2d}}
N^l \left(\frac{K}{N}\right)^b,
\end{equation}  
where $m_2, \cdots, m_{2d}$ indicate the number of intersections of $2k$
bonds ($k=1, \cdots,d$) forming a given tangle of $b$ bonds and $l$
loops.  An intersection of $2k$ bonds carries the
weight\cite{GerberFisher}
\begin{equation} 
\label{vw}
  \nu_{2k} = \frac{N^k \Gamma(N/2)}{2^k \Gamma(k+N/2)} = \frac{N^k}{N
    (N+2) \cdots (N+2k-2)}.
\end{equation} 
Due to the rescaling of the bond fugacity, $\nu_2=1$ for all $N$. In the
limit $N \to 0$, only the weight $\nu_2$ survives as $\nu_{2k} \to 0$
for $k>1$, implying that even on a hypercubic lattice, intersections are
completely suppressed and the high-temperature graphs reduce to
self-avoiding walks.

In the opposite limit, $N \to \infty$, all the vertex weights become
unity $\nu_{2k} \to 1$, and the partition function (\ref{Zsquare}) takes
the noninteracting form (\ref{phantom}) after undoing the rescaling of
the bond fugacity.  The only difference with the noninteracting theory
is that for a given number of set bonds $b$, only the graph with the
highest number of loops needs to be included as it dominates all other
graphs.  This limit corresponds to the so-called spherical
model.\cite{Stanley}

\begin{figure}
\begin{center}
\includegraphics[width=.8\textwidth]{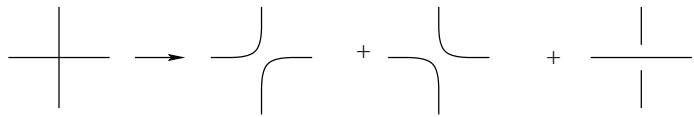}
\end{center}
\caption{A vertex with four bonds $(k=2)$ resolved into three possible
routings of the paths.
  \label{fig:crossing}}
\end{figure}

\begin{figure}
\begin{center}
\includegraphics[width=.6\textwidth]{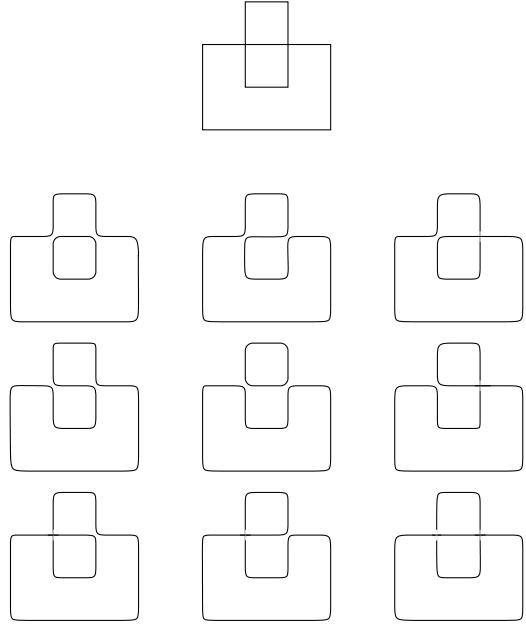}
\end{center}
\caption{A graph (top panel) involving two vertices with four bonds
$(k=2)$ is resolved into 9 different loop configurations according to
the rule displayed in Fig.~\ref{fig:crossing}.
  \label{fig:graph}}
\end{figure}

The loop representation (\ref{Zsquare}), featuring colorless loops,
obtains by resolving a vertex where $2k$ bonds of an high-temperature
graph meet into all possible routings of the paths (see
Figs.~\ref{fig:crossing} and \ref{fig:graph}).
The number of routings equals the number of ways $2k$ objects can be
divided into $k$ distinct pairs, i.e., $(2k-1)!!$.  Each of the routings
carries the same weight.  A given high-temperature graph is thus
resolved into loop configurations, and the partition function
(\ref{Zsquare}) describes a loop gas\index{loop!gas}, as does
Eq.~(\ref{Zhoneycomb}).  The loops physically represent worldlines that
can now intersect.  Although loops on a honeycomb lattice cannot do so,
for both the honeycomb and square lattice the same large-scale behavior
is obtained.  This implies that the operator which introduces
intersections is an irrelevant perturbation at the critical
point\index{critical!point|(}.\cite{Jacobsen}

It is worth pointing out that in the high-temperature representation,
the observation that $N=-2$ describes phantom loops is far from obvious.
On a square lattice, for example, a graph of $b$ bonds and one vertex
where four bonds meet contributes the following term to the partition
function,
\begin{equation} 
\lim_{N\to -2}  \nu_4 N (N+2) \left(\frac{K}{N}\right)^b = (-2)^2
\left(-\frac{K}{2}\right)^b,
\end{equation} 
which is finite because the weight factor $\nu_4$ given in
Eq.~(\ref{vw}) diverges in the limit $N\to-2$.  The diagrams no longer
cancel three by three as they did in Sec.~\ref{sec:phi4}
\subsection{Ising Model}
\label{sec:IM}
As an explicit example, we consider the simplest O($N$) model, viz.\ the
Ising model\index{Ising model} ($N=1$) on a square lattice.  The model
is defined by the Hamiltonian
\begin{equation} 
H = - J \sum_{\langle n n'\rangle} \phi_n \phi_{n'},
\end{equation} 
where the spins can take only two values $\phi_n = \pm 1$, and the
interaction parameter $J$ is taken to be unity.  The standard partition
function reads
\begin{equation} 
Z = \mathrm{Tr} \, \mathrm{e}^{-\beta H} = \mathrm{Tr} \prod_{\langle n
  n'\rangle} \mathrm{e}^{\beta \phi_n \phi_{n'}},
\end{equation} 
with the trace Tr denoting the sum over all possible spin
configurations.  In the context of statistical physics, $\beta$ denotes
the inverse temperature $\beta = 1/k_{\rm B} T$, with $k_{\rm B}$
Boltzmann's constant.  In the present context, we regard $\beta$ as
coupling constant.  Because of the identity
\begin{equation} 
\mathrm{e}^{\beta \phi_n \phi_{n'}} = \cosh \beta \left(1 + K \phi_n
\phi_{n'}\right),
\end{equation} 
with $K=\tanh\beta$, the standard representation can be rewritten in the
truncated form (\ref{ZS}) as:
\begin{equation} 
  Z = (\cosh \beta)^{2 N} \, \mathrm{Tr} \prod_{\langle n n'\rangle}  
 \left(1 + K \phi_n \phi_{n'}\right)
\end{equation}
up to an irrelevant prefactor.  Here, $2N$ denotes the total number of
bonds on a square lattice with $N$ sites and periodic boundary
conditions.  The right hand contains a total of $2N$ factors, one for
each bond, and therefore a total of $2^{2N}$ terms when multiplied out.
Each term can be visualized by a graph along the bonds on the lattice,
with each bond representing a pair of adjacent spins. Because of the
trace in the definition of the partition function, a graph having a
loose end does not contribute to $Z$.  Indeed, let the spin at the loose
end be $\phi_n$ then the sum over its possible configurations yields
zero, $\sum_{\phi_n=\pm 1} \phi_n =0$.  As a result, only
\textit{closed} graphs contribute.  Similarly, only those closed graphs
with an even number of bonds, i.e., two or four for a square lattice,
connecting each vertex of the graph yield nonzero contributions.  Such a
vertex contributes a factor of two as $\sum_{\phi_n=\pm 1} \phi_n^m =2$
for $m$ even.  The high-temperature representation of the Ising
model\index{Ising model} therefore involves a sum over all possible
closed graphs that can be drawn on the lattice:
\begin{equation} 
\label{HTIsing}
Z = (\cosh \beta)^{2N} 2^N
\sum_{\substack{\mathrm{closed} \\ \mathrm{graphs}}} K^b.
\end{equation}
A graph of $b$ bonds can be built from various disconnected
closed subgraphs.  The high-temperature representation is to be compared
with the loop representation (\ref{Zsquare}), which reduces to
\begin{equation} 
Z = \sum_\mathrm{loops} \left(\frac{1}{3}\right)^m  K^b,
\end{equation}  
for $N=1$ on a square lattice, with $m$ denoting the number of vertices
with four bonds.  Remembering that the loop representation is obtained
by resolving each of these vertices according to the rule given in
Fig.~\ref{fig:crossing}, we see that apart from an irrelevant prefactor
the two expressions coincide.  The Ising model\index{Ising model}
($N=1$) is special as the loop fugacity is unity, so that the
contribution of a closed graph depends only on the number of bonds it
occupies.

The correlation function of the Ising model\index{Ising model},
\begin{eqnarray} 
\left\langle \phi_n \phi_{n'} \right\rangle &=& \frac{1}{Z} \mathrm{Tr}
\, \phi_n \phi_{n'} \prod_{\langle m m'\rangle} \mathrm{e}^{\beta \phi_m
\phi_{m'}}, \nonumber \\ &=& \frac{1}{Z} (\cosh \beta)^{2N} \mathrm{Tr}
\, \phi_n \phi_{n'} \prod_{\langle m m'\rangle} \left(1 + K \phi_m
\phi_{m'}\right).
\end{eqnarray} 
can similarly be visualized by graphs on the lattice.  Each graph now
contains a subgraph connecting the lattice sites $n$ and $n'$.  All
other subgraphs must be closed and, except for the endpoints, all
vertices must be even as before.  In terms of these graphs,
\begin{equation} 
\left\langle \phi_n \phi_{n'} \right\rangle = \frac{1}{Z} (\cosh
\beta)^{2N} 2^N \sum_\mathrm{graphs} K^b.
\end{equation} 
\subsection{Summary}
In the spacetime approach to quantum field theory on a lattice, which is
geometric in nature, the partition function and the correlation function
are calculated in a hopping expansion\index{hopping!expansion}, thus
reducing the problem to counting paths on the lattice.  Physically, the
paths represent the worldlines traced out by the particles when
hopping\index{hopping} from one lattice site, representing a spacetime
cell, to an adjacent site.  The noninteracting lattice field theory
features phantom worldlines that can freely intersect and share bonds.
When the self-interaction is turned on, by taking the coupling constant
$\lambda>0$, intersections carry an energy penalty.  This has the effect
to suppress such configurations and as a result the particle
trajectories tend to avoid themselves.  In the limit of infinite
repulsion $\lambda \to \infty$, the $\phi^4$ lattice field
theory\index{lattice field theory|)} reduces to the O($N$) spin
model\index{O($N$) model}.  The hopping
expansion\index{hopping!expansion}, known in this context as the
high-temperature expansion, takes a particular simple form when the
truncated O($N$) model (\ref{ZS}) is implemented on a honeycomb lattice.
The worldlines are then by construction mutually and self-avoiding.
Since the large-scale behavior of this representative of the O($N$)
universality class on the honeycomb lattice is the same as for the
generic $\phi^4$ field theory\index{phi@$\phi^4$ theory} defined on a
square lattice, both multiply occupancy of bonds and intersections are
irrelevant.  At large scales, both loop gases\index{loop!gas} have the
same fractal structure.  Even when considering weak repulsion by taking
$\lambda$ small, after coarse-graining the fractal structure of the
loops appears the same as in the limit $\lambda\to\infty$.  This limit
therefore governs the large-scale behavior.

\section{Critical Properties}
\label{sec:CP}
In this section, the critical properties of the O($N$) universality
class are discussed from the perspective of worldlines.  In particular,
the fractal structure of the loop gas\index{loop!gas} featuring in the
spacetime approach to fluctuating fields close to the critical point is
studied.  It is shown how the critical
exponents\index{critical!exponents|(} characterizing the phase
transition can be obtained from the fractal structure of these
geometrical objects.  The discussion is geared after percolation theory,
which is purely geometric in nature.\cite{StauferAharony}
\subsection{Fractal Structure}
\label{sec:fractal}
Consider the O($N$) theory close to the point where it undergoes a
continuous phase transition to the ferromagnetic state, where the spins
prefer to point in one direction in spin space.  This phase transition
is associated with the spontaneous breaking of the O($N$) symmetry to
the subgroup O($N-1$) which includes rotations about the preferred spin
direction.  The logarithm of the partition function of the interacting
theory can in the vicinity of this critical point be written in a form
similar to the one found in Eq.~(\ref{Zell}) for the noninteracting
theory as
\begin{equation} 
\label{Zellp}
\ln Z/\Omega \sim \sum_b \ell_b.
\end{equation} 
The worldline loop distribution\index{loop!distribution} $\ell_b$
asymptotically splits into two factors as it did for the phantom loops
in Eq.~(\ref{ell}):
\begin{equation}
\label{ellb}
  \ell_b \sim b^{- \tau} \, \mathrm{e}^{- \theta b},
\end{equation} 
with $\tau$ an exponent characterizing the interacting theory.  The
value $\tau=d/2+1$ found for phantom loops is typical for a
noninteracting theory where the loops are simple closed Brownian
trajectories.  Formally, Eq.~(\ref{Zellp}) is the same as for
clusters\index{clusters|(} in percolation theory\index{percolation
theory} close to the percolation threshold.  As in that
context,\cite{StauferAharony} $\tau$ can be related to the fractal
dimension\index{fractal dimension} of the loops.  Consider the square
radius of gyration $R^2_b$ of a loop of $b$ steps
\begin{equation} 
R_b^2 = \frac{1}{b} \sum_{i=1}^b (n_i - \bar{n})^2 =
\frac{1}{2b^2} \sum_{i,j=1}^b (n_i - n_j)^2,
\end{equation} 
with $n_i$ denoting the lattice sites visited by the particle while
tracing out its worldline on the spacetime lattice, and $\bar{n} = (1/b)
\sum_{i=1}^b n_i$ the center of mass of the loop.  Asymptotically, the
average $\left\langle R^2_b \right\rangle$ scales with the number of
steps $b$ taken as
\begin{equation} 
\label{Hausdorff}
\left\langle R_b^2 \right\rangle \sim b^{2/D},
\end{equation} 
which defines the Hausdorff, or fractal dimension\index{fractal
dimension} $D$.  This average yields the typical linear size of a loop
of $b$ steps.  Specifically, such a loop is distributed over a volume of
typical linear size $\left\langle R_b^2 \right\rangle^{1/2}$.  Given the
definition of the loop distribution\index{loop!distribution} $\ell_b$ as
the average number (per lattice site) of loops of $b$ steps present, it
follows that $b \ell_b$ is the probability that a randomly chosen
lattice site belongs to such a loop, and
\begin{equation} 
b \ell_b \sim 1/\left\langle R^2_b \right\rangle^{d/2},
\end{equation} 
with $d$ the dimension of the spacetime box.  This leads to the relation
\begin{equation}
\label{tau} 
\tau = \frac{d}{D} + 1 ,
\end{equation}  
which is formally the same as in percolation theory.\cite{StauferAharony}

\subsection{Loop Proliferation\index{loop!proliferation}}
\label{sec:proliferation}
The Boltzmann factor in the loop distribution\index{loop!distribution}
(\ref{ellb}) exponentially suppresses long loops as long as the line
tension $\theta$ is finite.  Upon approaching the critical point, the
line tension vanishes as a power law,
\begin{equation} 
\label{theta}
\theta \sim (K_\mathrm{c}-K)^{1/\sigma},
\end{equation} 
with $\sigma$ a second exponent characterizing the loop
distribution\index{loop!distribution}.  When this happens, the loops can
grow arbitrarily long at no cost and the entire system
becomes\cite{Feynman55} ``pierced through and through with'' worldlines.
It is through this \textit{loop
proliferation}{loop!proliferation}\cite{Samuel} that a phase transition
manifests itself in the spacetime approach.  The phenomenon is
completely analogous to the sudden appearance at the percolation
threshold of a cluster spanning the infinite lattice in percolation
phenomena\index{critical!phenomena|)}.

In the field theoretic approach, the phase transition manifests itself
through a nonzero vacuum expectation value of the field $\phi^\alpha_n$,
where the component $\alpha$ indicates the preferred spin direction
spontaneously chosen by the system.  Since all possible directions are
equivalent, the system can choose any of these.  On a finite lattice,
the system can easily change its preferred direction.  This is in
particular the case for Monte Carlo\index{Monte Carlo} simulations using
a nonlocal update, such as the Swendsen-Wang,\cite{SwendsenWang} or
Wolff\cite{Wolff} algorithm, in which not individual spins are updated,
but entire clusters\index{clusters!spin}.  For this reason, the
following order parameter is frequently used in lattice
simulations:\cite{Binder}
\begin{equation} 
\mathcal{O} = \frac{1}{L^d} \left[\sum_\alpha \left(\sum_n \phi_n^\alpha
\right)^2 \right]^{1/2}.
\end{equation}   
With this choice of the order parameter, all possible directions are
treated equally. In the normal phase, the expectation value of this
operator vanishes.  At the critical point, a condensate forms and a
nonzero expectation value develops,
\begin{equation} 
 \left\langle \mathcal{O} \right\rangle \neq 0,
\end{equation} 
typical for the ferromagnetic state with the spins pointing in a
preferred direction in spin space.  Those particles tracing out long
worldlines all belong to the condensate.

The loop distribution\index{loop!distribution} (\ref{ellb}) is related
to the number of paths $z_b(a)$ of $b$ steps returning to a site
adjacent to the initial position through
\begin{equation}
  \label{ellz}
  \ell_b =  \frac{1}{b} z_b(a) K^b.
\end{equation}  
Because $z_b(a)$ refers to closed worldlines, the factor $1/b$ is
included to prevent overcounting as a given loop can be traced out
starting at any lattice site along that loop.  It should be noted that
in Eq.~(\ref{ellz}) rooted loops are considered, i.e., all loops run
through an arbitrary but fixed lattice site $n$.  The relation
(\ref{ellz}) implies that $z_b(a)$ scales as
\begin{equation} 
\label{zba}
z_b(a) K^b \sim b^{- d/D} \, \mathrm{e}^{- \theta b}.
\end{equation} 
Similarly, the number of paths $z_b$ of $b$ steps starting at $n$ and
ending at an arbitrary lattice site defined in Eq.~(\ref{zb}) scales as 
\begin{equation}
\label{zbs}
  z_b K^b \sim b^{\vartheta/D} {\rm e}^{- \theta b},
\end{equation}   
with $\vartheta$ a new exponent, characterizing the number of ways an
open path of $b$ steps can be embedded in the lattice.  It is worth
pointing out that in contrast to the loop exponents $\tau$ and $\sigma$,
this third exponent refers to open paths.  Its value is expected to be
positive because the number of possible rooted open paths with no
constraint on their endpoints increases with the number of steps $b$.
This is in contrast to the loop distribution\index{loop!distribution}
(\ref{ellb}), where the algebraic factor decreases with increasing $b$,
reflecting the increasing difficulty for a path to close.  According to
Eq.~(\ref{Pn}), the ratio of $z_b(n, n')$ and $z_b$ defines the
probability $P_b(n, n')$ for a particle to move from the lattice site
$n$ on the spacetime lattice to $n'$ in $b$ steps.  On general grounds,
it scales at criticality as:
\begin{equation}
  \label{P}
  P_b(n,n')  \sim   b^{-d/D} \, \mathsf{P} \left(|n-n'|/b^{1/D} \right),
\end{equation} 
with $\mathsf{P}$ a scaling function.  This generalizes the Gaussian
distribution (\ref{celebrate}) found for the noninteracting theory.
Consistency of the three scaling formulas (\ref{zba}), (\ref{zbs}), and
(\ref{P}), requires, as was first shown by McKenzie and
Moore\cite{McKenzieMoore} for self-avoiding random walks, that the
scaling function $\mathsf{P}(t)$ must vanish for $t \to 0$ as a power
law,
\begin{equation} 
\label{power}
\mathsf{P}(t) \approx t^\vartheta,
\end{equation} 
with the same exponent governing the \textit{asymptotic} behavior
(\ref{zbs}) of the number $z_b$ of open paths at the critical point
($\theta=0$).

\subsection{Critical Exponents}
\label{sec:ces}
As in percolation theory, the loop exponent $\sigma$ is related to the
exponent $\nu$, which specifies how the correlation length $\xi$
diverges at the critical point,
\begin{equation} 
\xi \sim |K - K_\mathrm{c}|^{-\nu}.
\end{equation} 
Physically, $\xi$ indicates the typical length scale in the system.  The
radius of gyration--a second typical length scale--can be written in
terms of $\xi$ as
\begin{equation} 
\left\langle R_b \right\rangle = \xi \, {\sf R}(b \theta),
\end{equation} 
where ${\sf R}$ is a scaling function and $\theta$ is the same parameter
as in the loop distribution\index{loop!distribution} (\ref{ellb}).  From
the asymptotic behavior (\ref{Hausdorff}), the divergence of the
correlation length, and the vanishing (\ref{theta}) of the line tension
$\theta$ as $K_\mathrm{c}$ is approached, the relation\cite{ht}
\begin{equation}
\label{nu} 
\nu = \frac{1}{\sigma D},
\end{equation} 
or
\begin{equation}
\label{nup} 
\nu = \frac{\tau-1}{d \sigma}
\end{equation} 
follows.  These relations are formally identical to those in
percolation theory.

The rest of the thermal critical exponents can be equally well expressed
in terms of the loop exponents and $\vartheta$.  To derive these
expressions, we write the correlation function in terms of $z_b(n,n')$
as on the honeycomb lattice [see Eq.~(\ref{GR})]
\begin{equation}
  \label{G}
\left\langle \phi^\alpha_n \phi^\beta_{n'} \right\rangle \sim
\frac{1}{2} \delta_{\alpha, \beta} \sum_b z_b(n,n') K^b = \frac{1}{2}
\delta_{\alpha, \beta} \sum_b z_b P_b (n,n') K^b .
\end{equation} 
When evaluated at criticality, where 
\begin{equation} 
\left\langle \phi^\alpha_n \phi^\beta_{n'} \right\rangle \sim
\delta_{\alpha, \beta}/|n-n'|^{d-2 + \eta},
\end{equation} 
Eq.~(\ref{G}) gives 
\begin{equation} 
\label{etavartheta}
\eta = 2 - D - \vartheta.
\end{equation}
This relation, with $\vartheta$ defined by Eq.~(\ref{power}), was
recently proposed for the three-dimensional XY\index{XY model}, i.e.,
O(2) model in Ref.~[\refcite{ProkofevSvistunov}].  This model describes
the superfluid phase transition\index{superfluid!phase transition} in
$^4$He.

Finally, using the definition of the magnetic susceptibility $\chi$,
\begin{equation} 
\chi = \sum_{n'} \left\langle \phi_n \cdot \phi_{n'}
\right\rangle \sim \sum_b z_b K^b, 
\end{equation} 
which diverges as $\chi \sim |K-K_\mathrm{c}|^{-\gamma}$, we find
\begin{equation} 
\gamma = \frac{1}{\sigma} \left(1 + \frac{\vartheta}{D} \right).
\end{equation} 
The explicit expressions given for $\nu,\eta$, and $\gamma$ satisfy
Fisher's scaling relation, $\gamma/\nu = 2-\eta$, and
\begin{equation} 
\frac{\gamma}{\nu} = D + \vartheta.
\end{equation} 

With $\nu = (\tau-1)/d \sigma$, Eq.~(\ref{Zell}) yields the scaling
relation $d \nu = 2 - \alpha$, where $\alpha$ determines the scaling
behavior of the free energy close to the critical point, 
\begin{equation} 
\ln Z/\Omega \sim |K-K_\mathrm{c}|^{2-\alpha}.  
\end{equation} 
The expressions for the other exponents follow by using the remaining
scaling relations.  We thus have shown that all the thermal critical
exponents are determined by the configurational entropy exponents for
closed and open particle trajectories, $\tau$ and $\vartheta$,
respectively, and by $\sigma$.

It is worth noting that the thermal critical exponents depend on only
\textit{two} independent variables, viz.\ $D+\vartheta$ and $\sigma D$.
Their significance in field theory is that they determine the anomalous
scaling dimensions of the $\varphi$ and $\varphi^2$ fields
\begin{equation} 
d_\varphi = \tfrac{1}{2} ( d - D-\vartheta) , \quad d_{\varphi^2} = d -
\sigma D,
\end{equation}
respectively.
\subsection{Self-Avoiding Random Walks}
\label{sec:saw}
Before applying the results of the previous section to the
two-dimensional O($N$) spin model\index{O($N$) model} with arbitrary $-2
\leq N \leq 2$, let us first verify that these general results are
consistent with the results known in the polymer limit $N \to
0$,\cite{deGennes} where, as we saw below Eq.~(\ref{vw}), the
high-temperature graphs reduce to self-avoiding walks.  Through the
exact enumeration and analysis of the number of self-avoiding loops on a
square lattice up to length 110, the value $\tau=5/2$ for the loop
exponent $\tau$ has been established to very high
precision.\cite{Jensen03} According to Eq.~(\ref{tau}), this corresponds
to the fractal dimension\index{fractal dimension} $D=4/3$--a value that
has been independently established to high precision in that same study
by determining the average square radius of gyration $\left\langle R_b^2
\right\rangle$ of the loops.  In a related study,\cite{Jensen04} where
the number $z_b$ of self-avoiding walks on a square lattice up to length
71 has been enumerated and analyzed, the value of the exponent
$\vartheta$, characterizing the open trajectories has been established.
Specifically, the value $\vartheta/D=11/32$ was found to high precision,
yielding $\vartheta = 11/24$.

In most studies on self-avoiding walks, the fractal
dimension\index{fractal dimension} of the walks is simply equated to the
inverse correlation length.  As follows from Eq.~(\ref{nu}), this
corresponds to setting $\sigma=1$.  A closer inspection of de Gennes'
original work on the $N \to 0$ limit reveals that there indeed
$\sigma=1$.

To sum up, the two configurational entropy exponents $\tau$ (loops) and
$\vartheta$ (open paths), and $\sigma$ for a self-avoiding walk in two
dimensions are given by
\begin{equation} 
  \tau = 5/2, \quad \vartheta = 11/24, \quad \sigma = 1.
\end{equation} 
With these values, all the critical exponents of the two-dimensional
O($N\to0$) model can be found through the relations of the previous
subsection and standard scaling relations. 
\subsection{O($N$) Models}
\label{sec:on}
The class of two-dimensional O($N$) models\index{O($N$) model} for $-2
\leq N \leq 2$ can be parametrized as
\begin{equation} 
N = - 2 \cos\left(\frac{\pi}{\bar\kappa}\right), 
\end{equation} 
with $\frac{1}{2} \le \bar\kappa \le 1$.  The fractal
dimension\index{fractal dimension} $D$ of the high-temperature loops is
known exactly\cite{DS88}
\begin{equation} 
D = 1 + \frac{\bar\kappa}{2},
\end{equation} 
leading to $\tau = (6 + \bar \kappa)/(2 + \bar \kappa)$.  The observable
whose scaling dimension is given by this fractal dimension\index{fractal
dimension} consists of two spins in different spin states placed at the
same site: $\phi_n^\alpha \phi_n^\beta$, with $\alpha \neq
\beta$.\cite{Nienhuis} It physically measures the tendency of spins to
align, as becomes intuitively clear by remembering that the spins
connected by a high-temperature graph are all in the same spin state.

While the fractal dimension\index{fractal dimension} of the
high-temperature loops is known exactly in two dimensions, the value of
the entropy exponent $\vartheta$ for open paths has to our knowledge not
been directly established for $N \neq 0$.  In two dimensions, it can be
established indirectly\cite{anomalous} by considering the Potts
model\index{Potts model}--a closely related spin model.

That model was reformulated by Fortuin and Kasteleyn\cite{FK} in a
purely geometric fashion in terms of suitably defined spin clusters.
The thermal phase transition is marked by a proliferation of these
clusters, which moreover have the thermal critical exponents encoded in
their fractal structure.  Percolation observables such as the average
cluster size and the percolation strength, defined as the probability
that a randomly chosen site belongs to the percolating cluster, serve as
improved estimators for the magnetic susceptibility and the
magnetization, respectively.  In other words, the thermal critical
exponents of the Potts model\index{Potts model} can be determined by
studying the fractal structure of these so-called Fortuin-Kasteleyn
clusters\index{clusters!Fortuin-Kasteleyn|(}, with the exponents defined
as in percolation theory.  In two dimensions, the fractal
dimension\index{fractal dimension} $D_\mathrm{C}$ of the
Fortuin-Kasteleyn clusters is known exactly\cite{Coniglio1989}
\begin{equation} 
  D_\mathrm{C} = 1 + \frac{\bar\kappa}{2} + \frac{3}{8}
  \frac{1}{\bar\kappa},
\end{equation} 
where the $Q$-state Potts models\index{Potts model} are parametrized in
terms of the same parameter $\bar\kappa$ parametrizing the O($N$)
models\index{O($N$) model} as
\begin{equation} 
\sqrt{Q} = - 2 \cos(\pi \bar\kappa),
\end{equation}  
with $\frac{1}{2} \le \bar\kappa \le 1$, so that $0\leq Q\leq4$.  The
$Q\to 0$ Potts model\index{Potts model} ($\bar \kappa = \frac{1}{2}$)
describes standard percolation theory.  Surprisingly, the external
perimeters of the Fortuin-Kasteleyn clusters have the same fractal
dimension\index{fractal dimension} $D$ as the O($N$)
loops.\cite{Duplantier00} For example, the external perimeters of
clusters\index{clusters|)} in standard percolation theory have the
dimension $\frac{4}{3}$ as have self-avoiding random
walks\index{self-avoiding walks|)} ($\bar \kappa=\frac{2}{3}$).  It can
be shown that the entropy exponent $\vartheta$ for open trajectories and
the two Potts dimensions satisfy the sum rule\cite{anomalous}
\begin{equation} 
d =  2 D_\mathrm{C} - D - \vartheta, 
\end{equation} 
with $d=2$ the dimension of the spacetime box.  With $D$ and
$D_\mathrm{C}$ known exactly, the expression for $\vartheta$ follows.

Finally, the expression for the exponent $\sigma$ can be inferred from
the values $\sigma=1$ for $N\to 0$ ($\bar \kappa=\frac{2}{3}$) and
$\sigma=0$ for $N=2$ ($\bar \kappa=1$).  This last value reflects the
special status of the XY model\index{XY model}, undergoing a phase
transition of the Berezinskii-Kosterlitz-Thouless type characterized by
algebraic long-range order.\cite{Berezinskii,KT} Assuming, as is the
case for Fortuin-Kasteleyn clusters, that both $\tau$ and $\sigma$ have
a similar dependence on $\bar\kappa$, we deduce that $\sigma = 8
(1-\bar\kappa)/(2+\bar\kappa)$.

To sum up this subsection, the two configurational entropy exponents
$\tau$ (loops) and $\vartheta$ (open worldlines), and $\sigma$ for the
O($N$) model\index{O($N$) model} with $-2 \leq N \leq 2$ are given in
terms of $\bar\kappa$ by
\begin{equation} 
  \tau = \frac{6+ \bar\kappa}{2+\bar\kappa}, \quad \vartheta = - 1 +
  \frac{\bar\kappa}{2} + \frac{3}{4} \frac{1}{\bar\kappa}, \quad \sigma
  = 8 \frac{1-\bar\kappa}{2+\bar\kappa}.
\end{equation} 
Table~\ref{table:On} collects the resulting values for the various
two-dimensional O($N$) models\index{O($N$) model} together with the
corresponding thermal critical exponents.  It is worth pointing out that
for $N=-2$ indeed Gaussian exponents are obtained, but that the fractal
dimension\index{fractal dimension} of the high-temperature graphs is not
given by that of a Brownian random walk, which has $D=2$.  Also note
that with increasing $N$, the exponent $\vartheta$ decreases, while at
the same time the fractal dimension\index{fractal dimension} $D$
increases.  The ratio $\vartheta/D$ appearing in the scaling relation
(\ref{zbs}) thus decreases with increasing $N$, implying that the number
of possible rooted open paths with no constraint on their endpoints
increases less rapidly with the number of steps $b$ for larger $N$.
\begin{table}
  \tbl{Critical exponents together with the exponents characterizing the
high-temperature graphs (worldlines) of various two-dimensional O($N$)
models\index{O($N$) model}.}  {
\begin{tabular}{l|cc|ccccc|cccc}
    \hline \hline & & & & & & & \\[-.1cm]
    Model & $N$ & $\bar\kappa$ & $\alpha$ & $\beta$ & $\gamma$ & $\eta$
    & $\nu$ & $D$ & $\tau$ & $\vartheta$ & $\sigma$ \\[.1cm]
    \hline & & & & & & & & & &  \\[-.1cm]
    Gaussian & $-2$ & $\frac{1}{2}$ & $1$ & $0$ & $1$ & $0$ & $\frac{1}{2}$ &  $\frac{5}{4}$ & 
    $\frac{13}{5}$ &  $\frac{3}{4}$ & $\frac{8}{5}$\\[.1cm]
    SAW & $0$ & $\frac{2}{3}$ & $\frac{1}{2}$ & $\frac{5}{64}$ & 
    $\frac{43}{32}$ & $\frac{5}{24}$ & $\frac{3}{4}$  & $\frac{4}{3}$ &  $\frac{5}{2}$ &  $\frac{11}{24}$ &$1$ \\[.1cm]
    Ising & $1$ & $\frac{3}{4}$ 
    & $0$ & $\frac{1}{8}$ & $\frac{7}{4}$ &
    $\frac{1}{4}$ & $1$ &
    $\frac{11}{8}$ &  $\frac{27}{11}$ &  $\frac{3}{8}$ & $\frac{8}{11}$ \\[.1cm]
    XY & $2$ & $1$ &  $-\infty$ & $\infty$ & $\infty$ & $\frac{1}{4}$ & $\infty$ 
    & $\frac{3}{2}$ &  $\frac{7}{3}$ &  $\frac{1}{4}$ & $0$\\[.1cm] \hline \hline
  \end{tabular}
    \label{table:On} 
} 
\end{table}

\subsection{Summary}
It was shown that the worldlines or high-temperature graphs of the
O($N$) theory proliferate right at the thermal critical point, and that
the entire set of critical exponents can be retrieved from the fractal
structure of these line objects.  In this way, a purely geometric
description of the phase transition in these systems was arrived at.

\section{Monte Carlo Simulations}
\label{sec:mc}
We next discuss a Monte Carlo\index{Monte Carlo} study of the Ising
model\index{Ising model} on a square lattice we carried
out\cite{geoPotts} to support the above findings (for a summary of that
study, see Ref.~[\refcite{cccp}]).  We consider not the original spin
formulation of the model, but the high-temperature representation
detailed in Sec.~\ref{sec:IM} instead, and examine whether the fractal
structure of the high-temperature graphs indeed encodes the thermal
critical behavior.  We focus exclusively on the graph configurations and
use percolation observables to study the fractal properties of the
graphs.

Traditionally, high-temperature
expansions\index{high-temperature!expansion|)} are carried out exactly
up to a given order by enumerating all possible ways graphs up to that
order can be embedded in the lattice.\cite{GFCM} We take a different
approach and study the high-temperature graphs by means of Monte
Carlo\index{Monte Carlo} methods.  The Metropolis
algorithm\cite{Erkinger} we use involves a local update and for that
reason suffers from critical slowing down close the the transition
temperature.  We have intensionally chosen the local update as it
allows for a particular clean implementation of our ideas.  The price we
pay for this is that we cannot study too large spacetime boxes.  The
largest lattice we can reasonably study is of linear size $L=256$.

\subsection{Plaquette Update\index{plaquette update}}
\label{sec:plaq}
\begin{figure}
\begin{center}
\includegraphics[width=.3\textwidth]{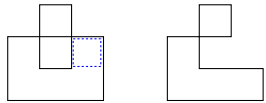}
\end{center}
\caption{Plaquette update\index{plaquette update}: An existing
high-temperature graph with the plaquette proposed for updating
indicated by the broken square (left panel) and the new graph after the
proposal is accepted (right panel).
\label{fig:update}}
\end{figure}

The update algorithm we use in our purely geometric approach of the
Ising model\index{Ising model} directly generates the high-temperature
graphs that contribute to the partition function.  Specifically, the
closed graphs are generated by means of a Metropolis \textit{plaquette}
algorithm.  A proposed plaquette update\index{plaquette update}
resulting in $b'$ occupied bonds is accepted with probability
\begin{equation} 
\label{Metro}
p_\mathrm{HT} = \left\{ \begin{array}{ll} K^{b'-b} & \quad \mbox{if}
    \;\; \quad b'>b \\ 1 & \quad \mbox{else} \end{array} \right. ,
\end{equation} 
where $b$ denotes the number of occupied bonds before the
update.\cite{Erkinger} In the entire temperature range $0 \leq
\beta \leq \infty$, $K=\tanh\beta \leq 1$.  Reflecting the Z$_2$
symmetry of the model, all bonds of an accepted plaquette are changed,
i.e., those that were occupied become unoccupied and \textit{vice versa}
(see Fig.~\ref{fig:update}).
With $b_\square$ denoting the number of bonds on the proposed plaquette already
occupied, it follows that
\begin{equation} 
\label{ll}
b' = b + 4 - 2 b_\square.
\end{equation}  
  
The prescription (\ref{Metro}) immediately follows from the
high-temperature representation (\ref{HTIsing}) of the Ising
model\index{Ising model}.  In equilibrium, the probability distribution
$P(G)$ for a given graph configuration $G$ involving $b$ occupied bonds
reads
\begin{equation} 
  P(G) = \frac{1}{Z} (\cosh \beta)^{2N} 2^N  K^b.
\end{equation} 
Such a configuration can be reached after $t+1$ iterations from the
configuration present after $t$ iterations in the following fashion
\begin{equation}
  P(G,t+1) = P(G,t) + \sum_{G'} \left[ P(G',t) W(G' \to G) -  
    P(G,t) W(G \to G') \right],
\end{equation} 
where $W(G \to G')$ is the probability for the system to move from the
graph configuration $G$ with $b$ occupied bonds to the graph
configuration $G'$ with $b'$ occupied bonds.  In equilibrium, $P(G,t+1)
= P(G,t) = P(G)$, and the system satisfies detailed balance
\begin{equation} 
  P(G') W(G' \to G)  =  P(G) W(G \to G'),
\end{equation} 
or
\begin{equation} 
\frac{W(G \to G')}{W(G' \to G)} = \frac{P(G')}{P(G)} = \frac{K^{b'}}{K^b} .
\end{equation} 
To maximize the acceptance rate of proposed updates, the largest of the
two transition probabilities $W(G \to G')$ and $W(G' \to G)$ appearing
in the ratio should be given the largest possible value, i.e., one.
Thus, if the number of bonds $b'$ in the proposed configuration is
larger than the number of bonds $b$ in the present configuration, so
that $K^{b'}/{K^b}<1$, then $W(G' \to G)=1$ and $W(G \to
G')=p_\mathrm{HT}$.  On the other hand, if $b'<b$, the proposed
configuration carries a larger weight than the present one, and will be
accepted unconditionally.

\begin{figure}
\begin{center}
\includegraphics[width=.43\textwidth]{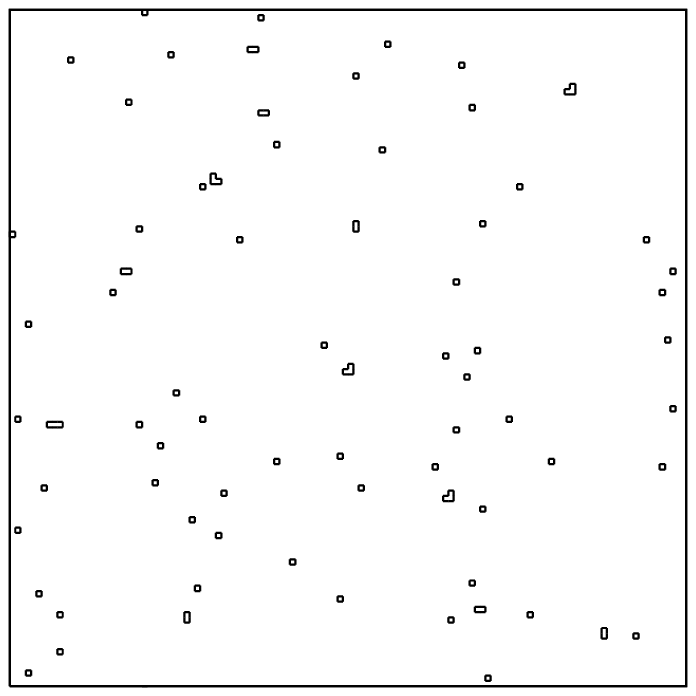} \hspace{.01\textwidth}
\includegraphics[width=.43\textwidth]{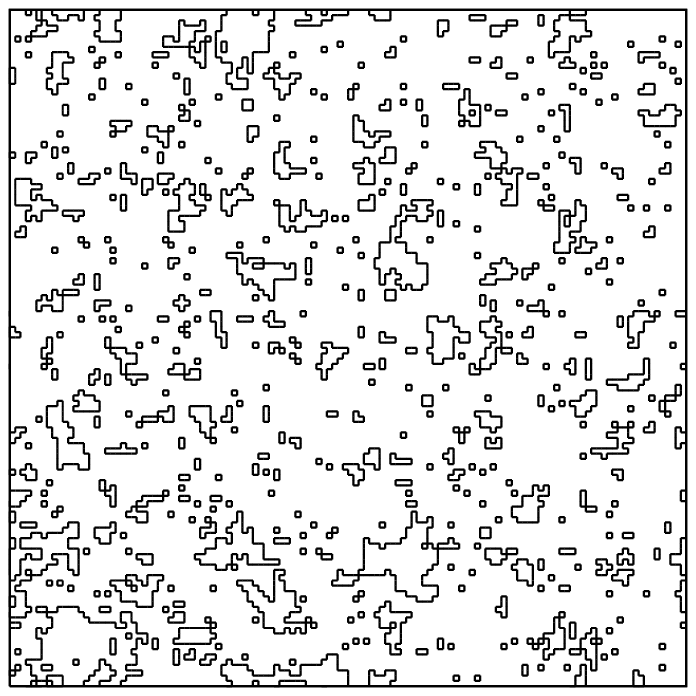} \\[.5cm]
\includegraphics[width=.89\textwidth]{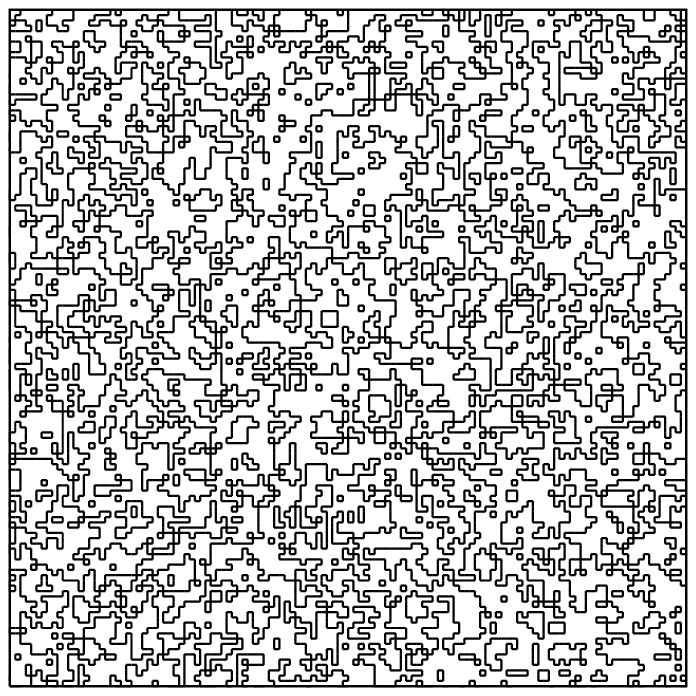}
\end{center}
\caption{Snapshots of high-temperature graphs of the Ising
model\index{Ising model} on a square lattice with linear size $L=128$
and periodic boundary conditions in the normal phase at $\beta=0.6
\beta_\mathrm{c}$ (top left panel), at the critical point $\beta =
\beta_\mathrm{c}$ (top right panel), and in the condensed phase at
$\beta=1.4 \beta_\mathrm{c}$ (bottom panel). Because of its richer
structure, the last snapshot is enlarged relative to the first two
snapshots.
\label{fig:snapshots}}
\end{figure}

By taking plaquettes, i.e., elementary loops on the spacetime lattice as
building blocks, the resulting high-temperature graphs are automatically
closed (see Fig.~\ref{fig:snapshots} for snapshots at three different
temperatures).

\subsection{Numerical Results}
\label{sec:num}

The data was collected on lattices varying in linear size from $L=16$ to
$256$ in $3.3 \times 10^5$ Monte Carlo\index{Monte Carlo} sweeps of the
lattice close to the critical point and $1.1 \times 10^5$ sweeps outside
the critical region.  About 10\% of the sweeps were used for
equilibration.  After each sweep, the resulting graph configuration was
analyzed.  Statistical errors were estimated by means of binning.

\begin{figure}
\begin{center}
\includegraphics[width=8.cm]{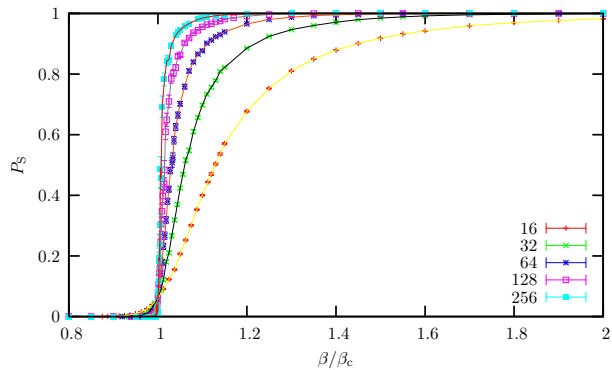} 
\end{center}
\caption{Probability $P_{\rm S}$ for the presence of a spanning graph as
function of the inverse temperature $\beta$ measured for lattice sizes
$L=16, 32, 64, 128, 256$.  Within the achieved accuracy, the curves
cross at the thermal critical point $\beta = \beta_{\rm c}$.
\label{fig:ps}}
\end{figure}

We first determine whether the high-temperature graphs proliferate
precisely at the inverse critical temperature $\beta_{\rm c} =
\ln(1+\sqrt{2})/2 = 0.440686\cdots$.  Remember that the inverse
temperature is related to the tuning parameter $K$ via $K=\tanh\beta$.
To this end, we measure the so-called spanning probability $P_{\rm S}$
as a function of $\beta$ for different lattice sizes.  Giving the
probability for the presence of a graph spanning the lattice, $P_{\rm
S}$ tends to zero for small $\beta$, while it tends to unity for large
$\beta$.  This observable has no scaling dimension and plays the role of
the Binder cumulant in standard thermodynamic studies.  When plotted as
a function of $\beta$, the curves $P_{\rm S}(\beta)$ obtained for
different lattice sizes should all intersect in one point.  Being
independent of the lattice size, this common point marks the
proliferation temperature of the \textit{infinite} system.  Within the
achieved accuracy, we observe that all the measured curves cross at the
thermal critical point, implying that the high-temperature graphs loose
their line tension $\theta$ and proliferate precisely at the Curie point
(see Fig.~\ref{fig:ps}).

\begin{figure}
\begin{center}
\includegraphics[width=8.cm]{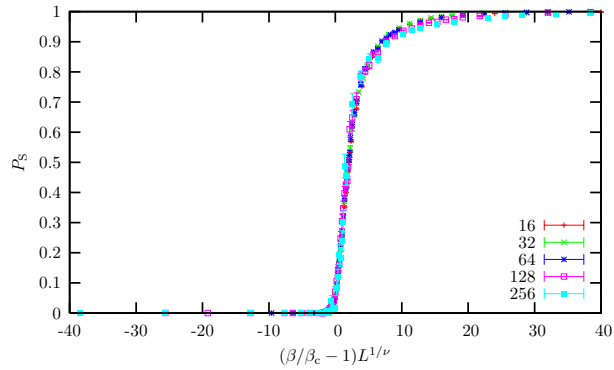}
\end{center}
\caption{Raw data of Fig.~\ref{fig:ps} replotted as a function of
$(\beta/\beta_{\rm c}-1)L^{1/\nu}$, with the Ising choice $\nu=1$.  The
data collapse is satisfactory over the entire temperature range.
  \label{fig:ps_collapse}}
\end{figure}

According to standard finite-size scaling\index{finite-size scaling},
this observable obeys the scaling law\cite{BinderHeermann}
\begin{equation} 
\label{FSSPS}
P_{\rm S}(\beta,L) = {\sf P}_{\rm S}(L/\xi),
\end{equation} 
where ${\sf P}_{\rm S}$ is a scaling function and $\xi$ the correlation
length whose divergence at criticality is governed by the exponent
$\nu$.  The data gathered on lattices of different sizes is therefore a
function not of $\beta$ and $L$ separately, but only of the ratio of the
lattice size and the correlation length.  All the data should therefore
collapse onto a single curve when plotted as a function of
$(\beta/\beta_{\rm c}-1)L^{1/\nu}$.  With $\nu$ given the Ising value
$\nu=1$, this is indeed what we find (see Fig.~\ref{fig:ps_collapse}).

\begin{figure}
\begin{center}
\includegraphics[width=8.cm]{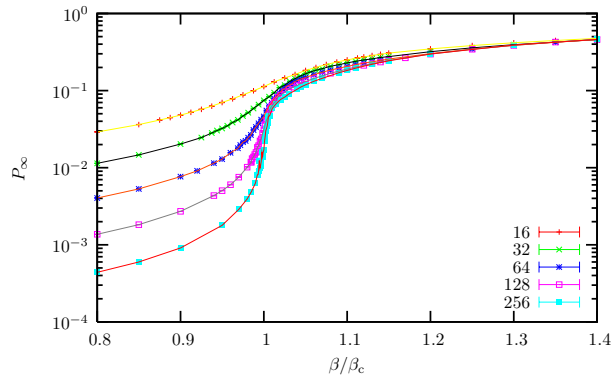}
\end{center}
\caption{Percolation strength as a function of the inverse temperature
$\beta$ measured on lattices of linear size $L=16, 32, 64, 128, 256$.
Note the logarithmic scale on the vertical axis.
  \label{fig:p8}}
\end{figure}

\begin{figure}
\begin{center}
\includegraphics[width=8.cm]{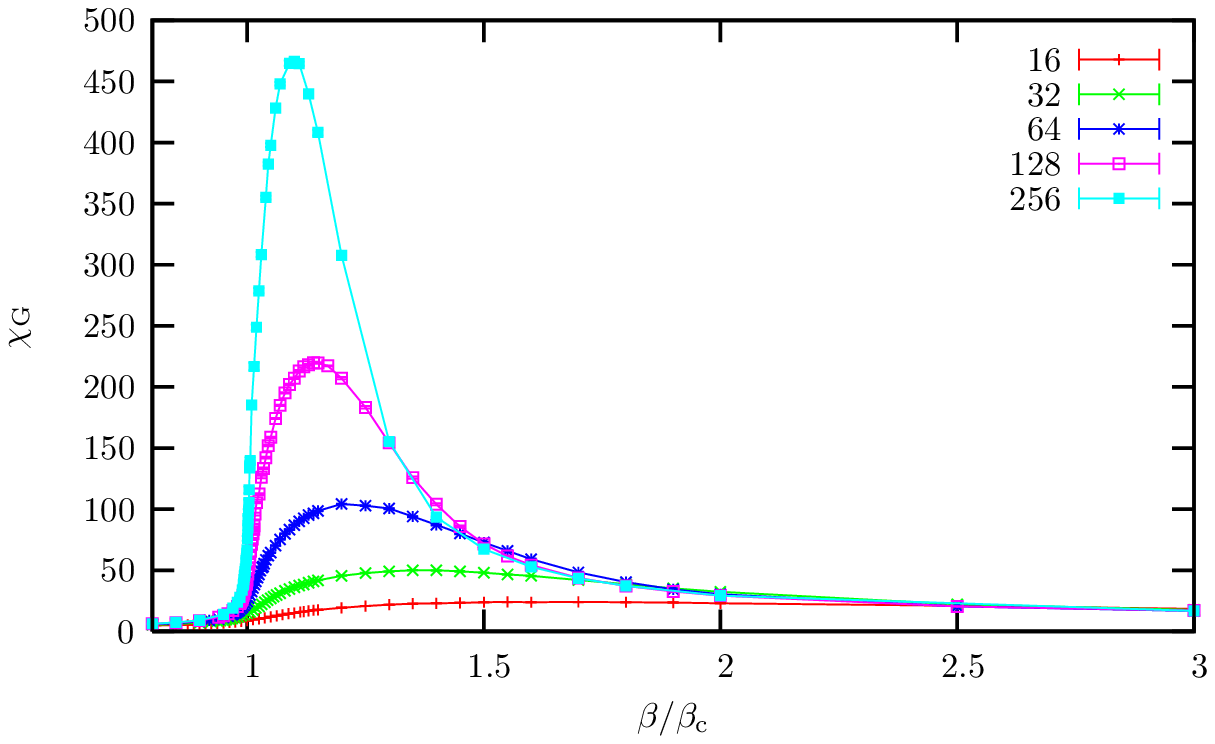}
\end{center}
\caption{Average cluster size $\chi_\mathrm{G}$, excluding the largest
cluster, as a function of the inverse temperature measured on lattices
of linear size $L=16, 32, 64, 128, 256$.
  \label{fig:chi}}
\end{figure}

The fractal dimension\index{fractal dimension} of the high-temperature
graphs is best determined by following standard percolation
theory\cite{StauferAharony} and measuring the percolation strength
$P^\mathrm{G}_\infty$, defined as the fraction of bonds in the largest
graph, and the average graph size $\chi_\mathrm{G}$ (see
Figs.~\ref{fig:chi} and \ref{fig:p8}).
The percolation strength, which is finite in the low-temperature phase
($\beta>\beta_\mathrm{c}$), vanishes as a power law when the percolation
threshold is approached,
\begin{equation} 
P^\mathrm{G}_\infty \sim (\beta-\beta_\mathrm{c})^{\beta_\mathrm{G}}.
\end{equation} 
At the same time, the average cluster size diverges,
\begin{equation} 
\chi_\mathrm{G} \sim |\beta-\beta_\mathrm{c}|^{-\gamma_\mathrm{G}}.
\end{equation}   
Here, $\beta_\mathrm{G}$ and $\gamma_\mathrm{G}$ are two percolation
exponents which should not be confused with the thermal critical
exponents.  The ratios $\beta_\mathrm{G}/\nu$ and
$\gamma_\mathrm{G}/\nu$ are expressed in terms of the entropy exponent
$\tau$ of the graph distribution as in percolation
theory\cite{StauferAharony}
\begin{equation} 
\frac{\beta_\mathrm{G}}{\nu} = d \frac{\tau-2}{\tau-1}, \quad
\frac{\gamma_\mathrm{G}}{\nu} = d \frac{3-\tau}{\tau-1},
\end{equation} 
with $d$ the dimensionality of the lattice.  According to
Table~\ref{table:On}, $\tau = 27/11 = 2.4546 \dots$ for the
two-dimensional Ising model\index{Ising model}.  Since $D=d/(\tau-1)$,
the fractal dimension\index{fractal dimension} is related to these
percolation exponents via
\begin{equation}  
\label{Dbg}
D = d - \frac{\beta_\mathrm{G}}{\nu} = \frac{1}{2} \left(d +
\frac{\gamma_\mathrm{G}}{\nu} \right),
\end{equation} 
again as in percolation theory.\cite{StauferAharony} Close to the
percolation threshold, the percolation strength and average graph size
obey the finite-size scaling\index{finite-size scaling} laws
\begin{equation} 
\label{finitess}
P^\mathrm{G}_\infty = L^{-\beta_\mathrm{G}/\nu} \, {\sf
P}^\mathrm{G}_\infty(L/\xi), \quad \chi_\mathrm{G} =
L^{\gamma_\mathrm{G}/\nu} \, {\sf X}_\mathrm{G} (L/\xi),
\end{equation} 
so that the exponent ratios and thus $D$ can be determined by
considering the system at criticality, where these scaling relations
imply an algebraic dependence on the system size $L$.
\begin{figure}
\begin{center}
\includegraphics[width=8.cm]{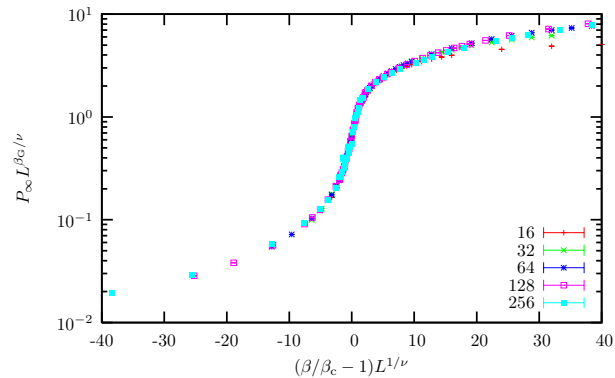}
\end{center}
\caption{Rescaled raw data of Fig.~\ref{fig:p8} plotted as a function of
$(\beta/\beta_{\rm c}-1)L^{1/\nu}$, with the Ising choice $\nu=1$, and
$\beta_\mathrm{G}=0.626$.  The data collapse is satisfactory in the
critical region as well as in the entire high-temperature phase
$\beta<\beta_\mathrm{c}$.  The flaring out of the data points for
$\beta>\beta_\mathrm{c}$ marks the end of the critical region.
  \label{fig:p8_collapse}}
\end{figure}

We numerically arrived at the estimates\cite{geoPotts} $\beta_{\rm G} =
0.626(7), \gamma_{\rm G} = 0.740(4)$, leading to $\sigma = 0.732(6),
\tau = 2.458(5)$ in perfect agreement with the exact values $\sigma =
8/11 = 0.7273 \dots,\tau = 27/11 = 2.4546 \dots$, and the predicted
fractal dimension\index{fractal dimension}\cite{DS88} $D = 11/8$ of the
high-temperature graphs.  The data were fitted over the range $L = 16 -
128$, using the least-squares Marquardt-Levenberg algorithm.  With the
obtained value for $\beta_{\rm G}$, a satisfactory collapse of the
percolation data is achieved (see Fig.~\ref{fig:p8_collapse}).  The
collapse of the average graph size data is less satisfactory. This is
not untypical for this type of observable.  A similar bad data collapse
we observed for the average size of geometrical (as opposed to
Fortuin-Kasteleyn\index{clusters!Fortuin-Kasteleyn|)}) spin
clusters\index{clusters!spin} in the two-dimensional Ising
model\index{Ising model}.
\subsection{Summary}
The case study of the Ising model\index{Ising model} on a square lattice
in the high-temperature
representation\index{high-temperature!representation|)} showed that the
fractal structure of the high-temperature graphs indeed encodes the
thermal critical behavior.  This representation therefore provides a
purely geometric and equivalent description of the original spin
formulation of the model.

\section{Further Applications}
\label{sec:appl}
The loop gas\index{loop!gas} approach is by no means limited to the
O($N$) theory.  The formalism is general and can be applied to a host of
other phase transitions, not necessarily involving the proliferation of
line objects.  Phase transitions involving the proliferation of surfaces
and hypersurfaces can be treaded similarly.  In this section, a few
further applications are briefly discussed.
\subsection{Higgs Model\index{Higgs!model|(}}
\label{sec:higgs}
A straightforward extension of the O($N$) theory follows by considering
electrically charged particles.  The continuum action describing the
charged system reads
\begin{equation} 
\label{Higgs}
S = \int \mathrm{d}^d x \left[ - \frac{1}{2} \left|
\left(\partial_\mu - \mathrm{i} e A_\mu \right) \varphi\right|^2 +
\frac{m^2}{2} |\varphi|^2 + \frac{g}{4!}  |\varphi|^4 + \frac{1}{4}
F^2_{\mu \nu}+ \frac{1}{2 \alpha} (\partial_\mu A_\mu)^2 \right]
\end{equation}
where $F_{\mu \nu} = \partial_\mu A_\nu - \partial_\nu A_\mu$ is the
electromagnetic field strength, with $A_\mu$ the vector potential in $d$
spacetime dimensions.  The scalar field $\varphi$ is now complex with
$N/2$ complex, i.e., $N$ real components, where $N$ is even.  The action
is obtained from the neutral theory by minimal coupling the scalar field
to the gauge field\index{gauge theory|(} $A_\mu$, with $e$ the electric
charge.  The Maxwell term provides the standard kinetic term for the
gauge field, while the last term with parameter $\alpha$ fixes a
Lorentz-invariant gauge.  The theory (\ref{Higgs}) is known as the
Abelian Higgs model in high energy and as the Ginzburg-Landau
theory\index{Ginzburg-Landau theory} in condensed matter physics.  The
partition function, whose general definition is given in Eq.~(\ref{Z}),
of the charged theory involves in addition to the functional
integral\index{functional integral} over the scalar field also one over
the gauge field, i.e.,
\begin{equation} 
\mathrm{Tr} = \int \mathrm{D} \varphi  \, \mathrm{D} A_\mu, 
\end{equation} 
where the functional measure $\int \mathrm{D} \varphi$ is the continuum
limit of the discrete measure (\ref{Tr}) defined on the spacetime
lattice, with an analogues definition for $\int \mathrm{D} A_\mu$.

To arrive at the spacetime description of this theory, we concentrate on
the coupling to the gauge field by considering only the first two terms
in the action,
\begin{equation} 
S_0 = \int \mathrm{d}^d x \left[ - \frac{1}{2} \left| \left(\partial_\mu -
\mathrm{i} e A_\mu \right) \varphi\right|^2 + \frac{m^2}{2} |\varphi|^2
\right].
\end{equation} 
This action describes otherwise free particles coupled to an
electromagnetic background field, specified by the gauge field $A_\mu$.
The corresponding partition function
\begin{equation}
\label{Zgauge} 
\ln Z_0 = \ln \mathrm{Det}^{-N/2}\left[ - \left(\partial_\mu- \mathrm{i}
e A_\mu \right)^2 + m^2  \right]
\end{equation} 
is readily written as a path integral\index{path integral} by
interpreting the operator $\left(\partial_\mu- \mathrm{i} e A_\mu
\right)^2 + m^2$ as the Hamiltonian operator of a nonrelativistic
charged particle of mass $M=\tfrac{1}{2}$ moving in a constant potential
$V=m^2$ in $d$ dimensions in the presence of an electromagnetic
background field.  The corresponding classical action is
\begin{equation} 
W_0 = \int_0^{s} \mathrm{d} s' \left\{\tfrac{1}{4} \dot{x}^2(s') -
\mathrm{i}  e \dot{x}_\mu A_\mu[x(s')] + m^2 \right\} .
\end{equation} 
According to the mnemonic discussed below Eq.~(\ref{lnZ0pi}), the
spacetime representation of the partition function (\ref{Zgauge}) is
given in terms of this action as
\begin{equation}
\label{lnZe} 
\ln Z_0 = \frac{N}{2} \Omega \int_0^\infty \frac{\mathrm{d}s}{s}  \oint
   \mathrm{D}x(s') \, \mathrm{e}^{-W_0}.
\end{equation} 
Since $N$ is even here, the simplest Higgs model corresponds to the
value $N=2$, which in the context of condensed matter physics provides
an effective description of ordinary
superconductivity\index{superconductivity|(}.

It follows from Eq.~(\ref{lnZe}) that a particle trajectory $\Gamma$
picks up an extra phase factor
\begin{equation} 
\label{U}
U(\Gamma) = \mathrm{e}^{\mathrm{i} e \int_0^s \mathrm{d} s'\,
\dot{x}_\mu A_\mu[x(s')]} = \mathrm{e}^{\mathrm{i} e \int_\Gamma
\mathrm{d} x_\mu A_\mu(x)}.
\end{equation} 
This phase factor results in a Biot-Savart type of
interaction\index{Biot-Savart interaction} between two line elements
$\mathrm{d} x_\mu$, $\mathrm{d} y_\nu$ of charged loops\cite{BMK} as can
be illustrated by considering the average of the phase factor (\ref{U})
\begin{eqnarray} 
\langle U(\Gamma) \rangle &\equiv& \int \mathrm{D} A_\mu \, U(\Gamma) \,
\exp \left\{- \int \mathrm{d}^d x \left[\frac{1}{4} F^2_{\mu \nu}+
\frac{1}{2 \alpha} (\partial_\mu A_\mu)^2 \right] \right\} \nonumber \\
&=& \exp\left[ -\frac{e^2}{2}\int_\Gamma \mathrm{d} x_\mu \, \mathrm{d}
y_\nu \, D_{\mu \nu}(x - y)\right],
\end{eqnarray} 
where $D_{\mu \nu }$ is the correlation function of the gauge field
\begin{equation}
D_{\mu \nu }(x) = \int \frac{\mathrm{d}^d k}{(2\pi)^d} \frac{1}{k^2}
\left[ \delta_{\mu \nu} - (1-\alpha) \frac{k_\mu k_\nu}{k^2} \right]
{\rm e}^{\mathrm{i} k \cdot x},
\end{equation} 
as can be read off from the action (\ref{Higgs}).  This long range
Biot-Savart interaction\index{Biot-Savart interaction} between line
elements of charged loops sets the Higgs model apart from the neutral
O($N$) theory, where the loops experience only a steric repulsion.

The $|\varphi|^4$ term in the action (\ref{Higgs}) can be treated as in
the O($N$) theory, leading to a steric repulsion for charged loops.
Pasting the pieces together, one arrives at the spacetime representation
of the partition function of the Higgs model\cite{StoneThomas,TS,Samuel}
\begin{eqnarray} 
\label{eloops}
Z = \int \mathrm{D} A_\mu && \!\! \exp\left\{- \int \mathrm{d}^d x \left[
\frac{1}{4} F^2_{\mu \nu}+ \frac{1}{2 \alpha} (\partial_\mu A_\mu)^2
\right] \right\} \nonumber \\ && \!\! \times \sum_{l=0}^{\infty} \frac{1}{l!}
\left(\frac{N}{2} \right)^l \prod_{r=1}^l \left[ \Omega \int_0^\infty
\frac{\mathrm{d} s_r}{s_r}  \oint \mathrm{D} x_r(s'_r) \right] \,
\mathrm{e}^{-W^{(l)}},
\end{eqnarray}  
with the $l$-particle action
\begin{eqnarray} \label{action}
W^{(l)} &=& \sum_{r=1}^l \int_0^{s_r} \mathrm{d} s_r'
\left\{\tfrac{1}{4} \dot{x}_r^2(s_r') + m^2 + \mathrm{i} e
\dot{x}_{r,\mu}(s'_r) \cdot A_\mu[x_r(s'_r)]\right\} \nonumber \\ && +
\frac{g}{6} \sum_{r,r'=1}^l \int_0^{s_r} \mathrm{d} s_r'
\int_0^{s_{r'}} \mathrm{d} s_{r'}' \, \delta \left[ x_r (s_r') -
x_{r'} (s_{r'}') \right] .
\end{eqnarray} 
The partition function (\ref{eloops}) represents a grand canonical
ensemble of fluctuating closed worldlines of arbitrary length and shape
traced out by electrically charged point particles.  Since the matter
field $\varphi$ is complex, the worldlines have an orientation now.

From this spacetime representation it follows that the logarithm of the
partition function can again be cast in the general form (\ref{Zellp}),
with the loop distribution\index{loop!distribution} now describing
charged loops.  Because of the additional long range Biot-Savart
interaction\index{Biot-Savart interaction}, the fractal structure of
these loops will be different from those featuring in the neutral
theory, which experience just a steric repulsion.  The physical picture
remains unchanged, however.  As in the neutral theory, only a few small
loops are present in the normal phase due to the finite line tension
$\theta$.  On entering the Higgs phase\index{Higgs!phase}, characterized
by a sign change in the mass term of the Higgs model, the line tension
vanishes, and the charged loops proliferate as they can become arbitrary
long at no cost.  The vacuum becomes filled with ``a spaghetti of
tangled loops''.\cite{TS}

\subsection{Bose-Einstein Condensation\index{Bose-Einstein condensation|(}}
\label{sec:bec}
The spacetime approach was originally formulated to describe
nonrelativistic quantum mechanics,\cite{Feynman48} which fundamentally
differs from its relativistic counterpart. A relativistic quantum
particle roams the space dimensions as well as the time dimension, so
that its worldlines are random walks in \textit{spacetime}.  Such
particle trajectories are parameterized by the Schwinger proper time
parameter, i.e., by their arc~length.  A nonrelativistic particle, on
the other hand, executes a random walk in space only.  Its time
coordinate is given by absolute Newtonian time, which elapses uniformly
forward.  In a nonrelativistic ensemble, all the particles execute a
random walk in space synchronously in time, with all the worldlines
parameterized by Newtonian time.
\begin{figure}
\begin{center}
\includegraphics[width=.4\textwidth]{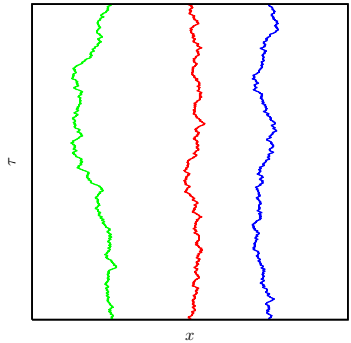}
\end{center}
\caption{Worldlines of three nonrelativistic particles on a square
lattice, representing one space and the imaginary time dimension, moving
synchronously from the bottom to the top of the lattice. The particles
execute a Brownian random walk in $x$-direction.  Periodic boundary
conditions are chosen in the imaginary time direction.
\label{fig:nrrw}}
\end{figure}
That is, the Schwinger proper time parameter gets replaced by Newtonian
time in the nonrelativistic limit.  At the technical level, this can be
seen by taking the nonrelativistic limit $c \to \infty$ in, for example,
the correlation function (\ref{chbar}) of the relativistic theory.  The
integral over $s$ can in this limit be approximated by the saddle
point\cite{Stephens}
\begin{equation} 
s = \tau/2m,
\end{equation}   
showing that the arc~length parameter $s$ of relativistic worldlines
indeed becomes replaced by Newtonian time $\tau$.  This difference has
profound implications.  Because Newtonian time always elapses forward,
nonrelativistic worldlines can no longer form simple loops on an
infinite lattice.  As a result, the number of particles present in the
system is fixed. On a finite lattice, worldlines still have the
possibility to close by wrapping around the lattice.

Figure \ref{fig:nrrw} shows an example of the motion in imaginary time
of three particles in one space dimension.  Up to now, the finite
lattice sizes were considered a mere approximation to systems of
infinite extent, in space as well as in time.  In this subsection, we
instead assume that the imaginary time dimension is finite.  With the
time variable $\tau$ restricted to the interval $0 \leq \tau \leq \hbar
\beta$,  and $\beta$ the inverse temperature $\beta = 1/k_{\rm B} T$,
one describes the system at finite temperature.  The zero-temperature
limit is recovered by letting the imaginary time dimension become
infinite.  For bosons, periodic boundary conditions are to be imposed in
the imaginary time dimension, implying that the configurations of an
ensemble of nonrelativistic bosons at imaginary time $0$ and at $\hbar
\beta$ are identical.  (In contrast to the rest of the paper, we
explicitly display factors of $\hbar$ in this subsection.)

The partition function $Z_l$ of such an ensemble of $l$ identical
nonrelativistic bosons can be expressed in terms of the probability
amplitude $G\left[\mathbf{x}(0),\mathbf{x}(\tau)\right]$ for a particle
to move from $\mathbf{x}(0)$ to $\mathbf{x}(\tau)$ in imaginary time
$\tau$ as:\cite{Matsubara}
\begin{equation} 
\label{Znr}
  Z_l = \sum_P \int \prod_{r=1}^l \mathrm{d}^d \mathbf{x}_r \,
      G\left[\mathbf{x}_1(0),\mathbf{x}_{P(1)}(\hbar \beta)\right]
      \cdots G\left[\mathbf{x}_l(0),\mathbf{x}_{P(l)}(\hbar
      \beta)\right] ,
\end{equation} 
where the final configuration $\{\mathbf{x}_{P(r)}(\hbar \beta)\}$ at
$\tau=\hbar \beta$ is a permutation $P$ of the initial configuration
$\{\mathbf{x}_r(0)\}$ at $\tau = 0$.  In contrast to the convention used
up to now, here, $d$ denotes the number of \textit{spatial} dimensions.
All possible permutations of the starting points are included as
indicated by the sum $\sum_P$.  In Feynman's spacetime approach, the
amplitude $G\left[\mathbf{x}(0),\mathbf{x}(\tau)\right]$ is written as a
sum over all possible particle trajectories connecting the two spatial
endpoints\cite{Feynman48}
\begin{equation} 
G\left[\mathbf{x}(0),\mathbf{x}(\hbar \beta)\right] = \int \mathrm{D}
      \mathbf{x}(\tau) \exp\left(-\frac{1}{\hbar} \int_0^{\hbar \beta}
      \mathrm{d} \tau \left\{\frac{m}{2} \dot{\mathbf{x}}^2(\tau) +
      V[\mathbf{x}(\tau)] \right\} \right),
\end{equation} 
where $V[\mathbf{x}(\tau)]$ is the potential experienced by the particle
along its worldline $\mathbf{x}(\tau)$.  Expressed as a path
integral\index{path integral}, the partition function thus
reads\cite{lambda}
\begin{equation} 
\label{Znrpi}
Z_l = \sum_P \int \prod_{r=1}^l \mathrm{D}
      \mathbf{x}_r(\tau) \, \exp \left(-\frac{1}{\hbar} \int_0^{\hbar
      \beta} \mathrm{d} \tau L^{(l)}\right)
\end{equation} 
with the $l$-particle nonrelativistic Lagrangian
\begin{equation} 
L^{(l)} = \sum_{r=1}^l \frac{m}{2} \dot{\mathbf{x}}_r^2(\tau) +
      \frac{1}{2} \sum_{r \neq r'=1}^l V\big[\left|\mathbf{x}_r(\tau) -
      \mathbf{x}_{r'}(\tau)\right|\big]
\end{equation} 
and pair potential $V(\mathbf{x}_r, \mathbf{x}_{r'}) = V(|\mathbf{x}_r -
\mathbf{x}_{r'}|)$.  Figure \ref{fig:nrrw} gives an example of a
configuration contributing to $Z_3$.  Since $\mathbf{x}_{P(1)}(\hbar
\beta) = \mathbf{x}_1(0)$, $\mathbf{x}_{P(2)}(\hbar \beta) =
\mathbf{x}_2(0)$, and $\mathbf{x}_{P(3)}(\hbar \beta) =
\mathbf{x}_3(0)$, the three particle trajectories form three separate
loops around the time axis.  This is typical for the high-temperature
phase of a nonrelativistic ensemble of bosons, where most particles
execute a random walk in space during the time interval $0 \leq \tau
\leq \hbar \beta$ that returns to its initial position.  It means that
the particles, being distinguishable, behave classically.

By mapping Feynman's spacetime approach onto the problem in classical
statistical mechanics of chains of beads connected by
springs\cite{ChanWol} to numerically evaluate the path
integral\index{path integral} with the help of the Metropolis Monte
Carlo\index{Monte Carlo} method, Ceperley and Pollock,\cite{CePo} and
others showed that a powerful computational tool emerges that can even
accurately describe a strongly interacting system like
superfluid\index{superfluid} $^4$He (for reviews, see
Refs.~[\refcite{SmithSinger,Ceperley}]).

Bose-Einstein condensation manifests itself in Feynman's spacetime
approach\index{spacetime approach|)} by the formation of so-called
\textit{cooperative exchange rings}, where individual particle
trajectories hook up to form longer loops.\cite{lambda}
\begin{figure}
\begin{center}
\includegraphics[width=.4\textwidth]{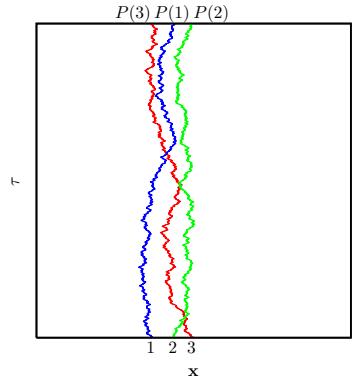}
\end{center}
\caption{Worldlines of three nonrelativistic particles on a hypercubic
lattice.  The vertical direction stands for the imaginary time
dimension, while the horizontal direction stands for all the space
dimensions.  The particles move synchronously from the bottom to the top
of the lattice, executing a Brownian random walk in \textit{space}.
Periodic boundary conditions are chosen in the time direction.  After
traversing the lattice, the particles are cyclically permuted $1 \to 2
\to 3 \to 1$.
\label{fig:nrrw_per}}
\end{figure}
An example is given in Fig.~\ref{fig:nrrw_per}.  Since
$\mathbf{x}_{P(1)}(\hbar \beta) = \mathbf{x}_2(0)$,
$\mathbf{x}_{P(2)}(\hbar \beta) = \mathbf{x}_3(0)$, and
$\mathbf{x}_{P(3)}(\hbar \beta) = \mathbf{x}_1(0)$, the three particle
trajectories now form a single loop, wrapping the time axis three times.
Generally, a particle in a composite exchange ring moves in imaginary
time along a trajectory that does not end at its own starting position,
but at that of another particle.  Whence, although the initial and final
configurations are identical, the particles in a composite ring are
cyclically permuted.  They thereby loose their identity and become
indistinguishable.  This is the essence of Bose-Einstein condensation,
which at finite temperature can occur in space dimensions larger than
two.

The connections between geometrical properties and critical exponents
spelled out in these notes apply equally well to Bose-Einstein
condensation.\cite{loops,pre01} Specifically, the logarithm of the
partition function (\ref{Znr}) takes close to the condensation
temperature again the form (\ref{Zellp}), with the worldline loop
distribution\index{loop!distribution} $\ell_b$ now denoting the number
density of worldlines wrapping around the time axis $b$ times.  In the
high-temperature phase, the loops have a finite line tension, so that
only small loops winding around the imaginary time axis only once or at
most a few times exist. Upon approaching the condensation temperature
from above, the line tension vanishes and loops with arbitrary large
winding numbers appear.\cite{Stone,loops} Since the winding number of a
given loop corresponds to the number of cyclically permuted particles in
an exchange ring in Feynman's theory, arbitrary large rings appear.  The
particles contained in large rings (long loops) are part of the
condensate.

For a free Bose gas in $2 < d < 4$ space dimensions, $\vartheta=0$ and
the fractal dimension\index{fractal dimension} of the particle
trajectories is that of a Brownian random walk, $D=2$, so that
$\tau=d/2+1$, while $\sigma=(d-2)/2$.\cite{pre01} Being related to the
thermal critical exponents as before (see Sec.~\ref{sec:ces}), these
exponents yield
\begin{equation} 
\label{ces}
\alpha = \frac{d - 4}{d - 2}, \quad \beta = \frac{1}{2}, \quad \gamma =
\frac{2}{d - 2}, \quad \nu = \frac{1}{d - 2}, \quad \eta = 0,
\end{equation} 
which are precisely the critical exponents of the spherical model in $d$
space dimensions.  This model corresponds to taking the limit $N \to
\infty$ in the O($N$) spin model\index{O($N$)
model}.\cite{GuntonBuckingham,DianaDimo} Despite being noninteracting, a
free Bose gas is not in the universality class of the Gaussian model.
The nontrivial exponents (\ref{ces}) derive from the constraint that the
total number of particles is fixed.  Without this constraint, which is
relevant when considering Bose-Einstein condensation in a free Bose gas
at constant pressure, Gaussian exponents
follow.\cite{GuntonBuckingham,DianaDimo} In four space dimensions,
corresponding to the upper critical dimension and above, i.e., $d \geq
4$, the exponents of the spherical and Gaussian models coincide.

The critical properties of Bose-Einstein condensation in an
\textit{interacting} nonrelativistic Bose gas are given by the $N=2$,
$\varphi^4$ universality class.  To state this more concretely, the
equilibrium thermal critical properties of Bose-Einstein condensation in
such an interacting system in three space dimensions, say, can be
computed from the theory (\ref{Scont}) with $d=3$, now considered a
\textit{classical theory} in three space dimensions.  The connection
between the nonrelativistic quantum theory and the classical O(2)
model\index{O($N$) model} is as follows.

Consider the square end-to-end vector $\left[\mathbf{x}_r(\tau) -
\mathbf{x}_r(0) \right]^2$ of the $r$th particle.  For a free Bose
gas, with the partition function (\ref{Znrpi}), its average is readily
calculated with the result in $d$ space dimensions
\begin{equation} 
\left\langle\left[\mathbf{x}_r(\tau) - \mathbf{x}_r(0) \right]^2
\right\rangle = d \frac{\hbar}{m} \tau.
\end{equation} 
This expression shows that the worldlines of free nonrelativistic
particles can be interpreted as trajectories traced out in
\textit{space} by a random walker taking steps of typical size
\begin{equation} 
a = \sqrt{\frac{d}{2 \pi}} \, \lambda_T
\end{equation} 
during each imaginary time interval $\tau=\hbar \beta$.  Here,
$\lambda_T$ is the de Broglie thermal wavelength
\begin{equation} 
\lambda_T \equiv \sqrt{\frac{2 \pi \hbar^2 \beta}{m}}.
\end{equation} 
The step size of the coarse-grained random walk is small when the
thermal wavelength is small, i.e., at high temperatures ($\beta$ small)
and for large particle masses.  These conditions correspond to the
classical limit.  Bose-Einstein condensation sets in when the step size
becomes on the order of the interparticle distance $1/n^{1/d}$.  Setting
$a=1/n^{1/d}$ leads to an estimate of the condensation temperature of a
free Bose gas
\begin{equation}
\label{T0d}
k_\mathrm{B} T_0 = \frac{2 \pi \hbar^2}{m} \left(
\frac{n}{\zeta(d/2)} \right)^{2/d},
\end{equation}
with $\zeta$ the Riemann zeta function
\begin{equation} 
\zeta(s) = \sum_{l=1}^\infty \frac{1}{l^s} ,
\end{equation} 
and where the particle number density $n$ is assumed to be given.  If
this density is small or if the mass of the particles is large, the
condensation temperature is low.

The inclusion of a self-interaction in the free (3+1)-dimensional
nonrelativistic quantum field theory changes the fractal structure of
the worldlines, as they acquire a steric repulsion now, but leaves the
basic picture unchanged.  Consider a grand canonical nonrelativistic
ensemble
\begin{equation} 
Z = \sum_{l=0}^\infty \frac{1}{l!} \, \mathrm{e}^{\beta \mu l} \, Z_l,
\end{equation} 
with $\mu$ the chemical potential and $Z_l$ given in Eq.~(\ref{Znrpi}).
The coarse-grained random walks of the interacting theory in
three-dimensional space can be identified with the high-temperature
graphs of the three-dimensional O(2) spin model\index{O($N$) model}.
Each set bond corresponds to a particle wrapping the time axis once.  On
a cubic lattice, the smallest closed graph involves four bonds, so that
translated back to the nonrelativistic quantum theory, exchange rings
smaller than four are not visible in the spin model.  In the
high-temperature phase of the spin model, only a few small
high-temperature graphs are present.  This reflects the fact that in the
nonrelativistic quantum theory most worldlines wrap the imaginary time
axis only once and that very few larger exchange rings are present at
high temperatures.  A large single closed high-temperature graph of $b$
bonds amounts in the nonrelativistic quantum theory to a large exchange
ring involving $b$ particles.  The proliferation of high-temperature
graphs in the spin model corresponds in the nonrelativistic quantum
theory to the proliferation of worldlines, wrapping arbitrary many times
around the imaginary time axis, and signals the onset of Bose-Einstein
condensation.
\subsection{Summary}
The formalism developed in these notes is shown to be easily adapted to
describe charged systems as well as Bose-Einstein
condensation\index{Bose-Einstein condensation|)}.

\section{Dual Theories\index{dual theory|(}}
\label{sec:dual}

As was first pointed out by Helfrich and M\"uller,\cite{HM} the
high-temperature graphs of O($N$) spin models\index{O($N$) model}
describe in addition to the particle trajectories at the same time a
second set of \textit{physical} lines.  In this section, we briefly
discuss examples in $d=2,3$ and 4, respectively.
\subsection{Peierls Domain Walls\index{domain walls}}
\label{sec:peierls}
Consider the Ising, i.e, the O(1) model\index{Ising model} on a square
lattice.  According to the famous Kramers-Wannier duality,\cite{KrWa}
the high-temperature graphs of the model form line-like Peierls
\textit{domain walls}\index{domain walls},\cite{Peierls} separating
geometrical spin clusters\index{clusters!spin} of opposite orientation on
the dual lattice.  Each bond in a high-temperature graph cuts a nearest
neighbor pair of anti-parallel spins on the dual lattice in two.  The
link connecting the two anti-parallel spins on the dual lattice is
perpendicular to the high-temperature bond on the original lattice.  On
an infinite lattice, the Kramers-Wannier duality implies that
observables calculated at an inverse temperature $\beta$ in the original
Ising model\index{Ising model} can be transcribed to those of the dual
model at an inverse temperature $\tilde\beta$.  The duality map
interchanges the low-temperature and high-temperature phases.  The
relation between the two temperatures is readily established by noting
that an occupied high-temperature bond represents a factor $K = \tanh
\beta$.  The anti-parallel spin pair on the dual lattice, that is cut by
the high-temperature bond, carries a Boltzmann weight $\exp(-2\tilde
\beta)$, so that\cite{KrWa}
\begin{equation} 
\label{bb}
\tanh \beta = {\rm e}^{-2\tilde \beta},
\end{equation} 
or $\sinh 2 \beta = 1/\sinh 2 \tilde\beta$.  The critical temperature
$\beta_{\rm c}$ follows from this relation by setting $\beta =
\tilde \beta$, with the result
\begin{equation} 
\beta_{\rm c} = \ln(1+\sqrt{2})/2 = 0.44068679\cdots .
\end{equation} 

By the Kramers-Wannier duality, the fractal dimension\index{fractal
dimension} of the line-like Peierls domain walls\index{domain walls} at
criticality coincides with the one of the high-temperature graphs, i.e.,
$D=\frac{11}{8}$.
\subsection{Vortex Lines\index{vortices|(}}
\label{sec:vortex}
We next consider the three-dimensional O(2) model\index{O($N$) model}.
The fractal dimension\index{fractal dimension} of the corresponding
high-temperature graphs has been estimated in a recent Monte
Carlo\index{Monte Carlo} study\cite{ProkofevSvistunov} as $D =
1.7655(20)$ at the critical point.  By a duality
map,\cite{BMK,Peskin,TS} these high-temperature graphs describe at the
same time the magnetic vortex lines of the three-dimensional Higgs model
(\ref{Higgs}).  Vortices constitute the topological defects of the Higgs
model and are line-like in three dimensions.  The duality map again
interchanges the low-temperature and high-temperature phases of the two
models.  Given the values\cite{3XY} $\eta=0.0380(4)$ and
$\nu=0.67155(27)$ of the O(2) model\index{XY model} obtained through
Monte Carlo\index{Monte Carlo} simulations, the relations
(\ref{etavartheta}) and (\ref{nu}) lead to the values
\begin{equation} 
\label{exps3xy}
\tau = 6.2965(60) , \quad \vartheta = 0.1965(20), \quad \sigma =
0.8434(13) ,
\end{equation} 
for the configurational entropy exponents of closed and open magnetic
vortices, $\tau$ and $\vartheta$, respectively, and for the exponent
$\sigma$, characterizing the vanishing of the line tension when the
critical point is approached.  The estimate for $\vartheta$ was first
given in Ref.~[\refcite{ProkofevSvistunov}].

From the perspective of vortices, the phase diagram of the
three-dimensional Higgs model is as follows.  For convenience, we take
the model as effective theory of ordinary type-II superconductors in
three space dimensions.  Whereas in the superconducting phase only a few
small vortex loops are present, the density and size of the loops
increase when the temperature rises.  Loops of all sizes appear at the
critical point, and the vacuum becomes filled with vortex loops.  The
disordering effect of the proliferating vortices destroys the
superconducting phase and drives the phase transition to the normal
phase.  Because of the Meissner effect, the magnetic interaction is
screened in the superconducting phase and vortex line elements
experience only a short range interaction which is well approximated by
a steric repulsion.  This is precisely the type of interaction
experienced by the high-temperature graphs of the O(2)
model\index{O($N$) model}.

The three-dimensional O(2) model\index{O($N$) model} also possesses
vortex lines as topological solutions.  In contrast to those in a
superconductor, the vortices in a superfluid\index{superfluid}
experience a long range interaction, mediated by the gapless Goldstone
mode associated with the spontaneous breaking of the O(2) symmetry in
the $|\varphi|^4$ theory\index{phi@$\phi^4$ theory}.  The model whose
high-temperature graphs describe these vortices is the Higgs model, as
follows again from the duality map.\cite{BMK,Peskin,TS} The long range
interaction experienced by the high-temperature graphs of the Higgs
model is the Biot-Savart interaction\index{Biot-Savart interaction}
discussed in Sec.~\ref{sec:higgs}.  The duality between the O(2) and the
Higgs models implies that the particle trajectories of one model appear
as vortex lines in the other.\cite{TS}

It was first suggested by Onsager\cite{Onsager} that proliferating
vortex loops\index{loop!proliferation} drive the superfluid phase
transition\index{superfluid!phase transition} in liquid $^4$He.  He
envisaged that as the critical temperature is approached from below, the
vortex loops proliferate{loop!proliferation} and thereby disorder the
superfluid\index{superfluid} state, causing the system to revert to the
normal state.  The precise nature of the three-dimensional superfluid
phase transition was recently investigated from this perspective in a
recent high-precision Monte Carlo\index{Monte Carlo}
study.\cite{vortexline} One of the observables considered is the total
vortex line density $v$.  By means of standard finite-size scaling
analysis of the corresponding susceptibility
\begin{equation} 
\chi = L^3(\langle v^2 \rangle - \langle v \rangle^2),
\end{equation}  
the inverse critical temperature $\beta_\mathrm{c}$ was estimated and
shown to be consistent with the estimate of a previous study directly in
terms of the field variables.\cite{bittnerjanke}  However, when
percolation observables were considered, such as the probability for the
presence of a vortex loop spanning the lattice, slight but statistically
significant deviations from $\beta_\mathrm{c}$ were found.  For all
observables considered, the vortex proliferation threshold
$\beta_\mathrm{p}$ is larger than $\beta_\mathrm{c}$.  That is, from
these observables one would conclude that the vortices proliferate too
early at a temperature below the critical one.  From the duality map
alluded to above, the vortex proliferation threshold is expected to
coincide with the critical point.  It is not clear at the moment, whether
this discrepancy is physical, or an artifact of the percolation
observables used.  With the percolation threshold taken as an
adjustable parameter, reasonable estimates were obtained from
percolation observables for the critical exponents $\nu$ and $\beta$,
consistent with those of the XY model\index{XY model}.\cite{vortexline}

Even in systems without a thermodynamic phase transition, the notion of
vortex proliferation can be useful in understanding the phase structure.
An example is provided by the three-dimensional Abelian Higgs lattice
model with \textit{compact} gauge field\index{gauge theory!compact} (see
below).\cite{compact} In addition to vortices, the compact model also
features magnetic monopoles\index{magnetic monopoles} as topological
defects, which are point-like in three dimensions.  In the presence of
monopoles, vortex lines no longer need to be closed as they can
originate at a monopole and terminate at an antimonopole.  When the
vortex line tension is finite, monopoles and antimonopoles are tightly
bound in pairs.  The part of phase diagram where this is the case is
called the Higgs region, which corresponds to the superconducting phase
in the noncompact theory.  When the vortex line tension vanishes,
monopoles and antimonopoles are no longer bound in pairs, and instead
form a plasma.  In the part of the phase diagram where this is the case,
the system exhibits charge confinement\index{confinement}.  It is well
established that one can move from the Higgs region into the confining
region without encountering thermodynamic
singularities.\cite{FradkinShenker} The susceptibility data for various
observables define, however, a precisely located phase boundary.  More
specifically, for sufficiently large lattices, the maxima of the
susceptibilities at the phase boundary do not show any finite-size
scaling, and the susceptibility data obtained on different lattice sizes
collapse onto single curves without rescaling, indicating that the
infinite-volume limit is reached.  In Ref.~[\refcite{compact}], it was
argued that this phase boundary marks the vortex proliferation
threshold.  A well-defined and precisely located phase boundary across
which geometrical objects proliferate, yet thermodynamic quantities
remain nonsingular has become known as a \textit{Kert\'esz line}.  Such
a line was first introduced in the context of the Ising
model\index{Ising model|)} in the presence of an applied magnetic
field.\cite{kertesz}
\subsection{Monopole Loops\index{monopole loops}}
\label{sec:monopole}
As a last example, we consider the four-dimensional \textit{pure
compact} U($1$) lattice gauge theory\index{gauge theory!compact}
described by the Wilson action\cite{Wilson}
\begin{equation}
\label{Sg} 
  S_g = \beta \sum_{x,\mu<\nu} \left[ 1-\cos \theta_{x, \mu \nu}
  \right].
\end{equation} 
Here, $\beta$ is the inverse gauge coupling, the sum extends over all
lattice sites $x$ and lattice directions $\mu$, and $\theta_{x, \mu
\nu}$ denotes the plaquette variable
\begin{equation} 
\theta_{x, \mu \nu} \equiv \Delta_\mu \theta_{x, \nu} -
\Delta_\nu\theta_{x, \mu} ,
\end{equation} 
with the lattice difference operator $\Delta_\nu \theta_{x,\mu} \equiv
\theta_{x+a\hat{\nu},\mu} - \theta_{x,\mu}$ and the compact variable
$\theta_{x, \mu} \in[-\pi,\pi)$, living on the link connecting the
lattice site $x$ with $x+a\hat{\mu}$.  The link variable is related to
the continuum gauge field $A_\mu(x)$ via
\begin{equation} 
\theta_{x, \mu} = e a A_\mu(x).
\end{equation} 
The lattice action (\ref{Sg}) reduces to the ordinary Maxwell action in
the continuum limit $a \to 0$, provided one sets
\begin{equation} 
\beta = \frac{1}{a^{4-d} e^2}.
\end{equation} 
As for the compact Higgs model\index{gauge theory!compact}, the pure
compact gauge theory possesses point-like monopoles in three dimensions.
In four dimensions, these defects become line-like.\cite{Polyakov_mon}
The monopole loops\index{monopole loops} experience in addition to a
steric repulsion also a long range Biot-Savart
interaction\index{Biot-Savart interaction}.  The field theory in which
these monopole lines appear as particle trajectories (high-temperature
graphs) is the four-dimensional noncompact Higgs model
(\ref{Higgs}).\cite{BMK,Peskin,StoneThomas} That is, the pure compact
U(1) lattice gauge theory\index{gauge theory|)} is dual to the
noncompact Higgs model.  These models possess two phases.  The pure
compact U(1) theory\index{gauge theory!compact} has a confining
phase\index{confinement} at strong coupling (small $\beta$),
corresponding to the superconducting\index{superconductivity|)}. phase
of the Higgs model\index{Higgs!model|)}., and a Coulomb phase at weak
coupling (large $\beta$) characterized by a massless photon.  Monte
Carlo\index{Monte Carlo} simulations\cite{DeGrandToussaint} show that in
the Coulomb phase only a few small monopole loops\index{monopole loops}
are present.  With increasing coupling constant (decreasing $\beta$),
the density and size of the loops increase.  At the critical
point\index{critical!point|)} very close to $\beta_\mathrm{c} = 1.0$,
the monopole loops\index{monopole loops}
proliferate\index{loop!proliferation} and thereby disorder the
system.\cite{Gupta:1985rz,Baig:1994ib} The Coulomb phase gives way to
the strong-coupling confining phase\index{confinement}, where the vacuum
is filled with a spaghetti of tangled monopole loops. It is worth
emphasizing that the monopole loops drive the phase transition, i.e.,
the monopole loops\index{monopole loops} proliferate right at the
confinement phase
transition\index{confinement}.\cite{Gupta:1985rz,Baig:1994ib} In showing
this, percolation observables\index{percolation theory|)} of the type
discussed here have been used.\cite{Hands:1989cg} The order of the phase
transition is not an established fact.  In case it is continuous, the
relations between the fractal structure of loops and the critical
exponents\index{critical!exponents|)} laid out in these notes also apply
here.
\subsection{Summary}
Besides picturing the worldlines\index{worldlines|)} of the particles
described by the field theory under consideration, the high-temperature
graphs\index{high-temperature!graphs|)} can have a second interpretation
as topological line defects, such as Peierls domains walls\index{domain
walls} $(d=2)$, vortices\index{vortices|)} $(d=3)$, or monopole
lines\index{magnetic monopoles} $(d=4)$.  The field theory in which such
line-like configurations appear as topological defects is said to be
dual\index{dual theory|)} to the original one.  Both field theories
describe the same physical lines, once interpreted as particle
trajectories\index{particle trajectories|)}, once as line defects.

\vspace{.5cm}

\noindent
\textbf{Acknowledgments} \\ One of us (A.S.) kindly thanks
Prof.~Y. Holovatch for the invitation to present one of the Ising
Lectures-2004 at the Institute for Condensed Matter Physics of the
National Academy of Sciences, Lviv, Ukraine.  He gratefully acknowledges
the hospitality and financial support provided by the Institute.  

This work is partially supported by the DFG grant No. JA 483/17-3 and by
the German-Israel Foundation (GIF) under grant No.\ I-653-181.14/1999.

\printindex

\end{document}